\documentclass[10pt,onecolumn,amsmath,amssymb, aps,prl,floatfix,superscriptaddress]{revtex4-2}

\newcommand{\env}[1]{\texttt{#1}}
\def \Pe {\mathrm{Pe}}

\usepackage{graphicx}
\usepackage{docs}%
\usepackage{nameref}
\usepackage[colorlinks=true,linkcolor=blue]{hyperref}
\usepackage{xcolor}
\usepackage{ulem}

\begin{document}

\title{Modeling complex motility patterns for autophoretic microswimmers}

\author{Anupriya Dutta Roy}%
\email{ph20resch11020@iith.ac.in}
\affiliation{
 Department of Physics,\\
Indian Institute of Technology Hyderabad,
Kandi, Sangareddy 502285, India }
\author{Smita S. Sontakke}%
\affiliation{
 Department of Mechanical and Aerospace Engineering,\\
Indian Institute of Technology Hyderabad,
Kandi, Sangareddy 502285, India }
\author{Arvind Kumar}%
\affiliation{
 Center for Interdisciplinary Programs,\\
Indian Institute of Technology Hyderabad,
Kandi, Sangareddy 502285, India }
\author{Ranabir Dey}%
\email{ranabir@mae.iith.ac.in}
\affiliation{ 
Department of Mechanical and Aerospace Engineering,\\
Indian Institute of Technology Hyderabad,
Kandi, Sangareddy 502285, India }
\author{Anupam Gupta}%
\email{agupta@phy.iith.ac.in}
\affiliation{
 Department of Physics,\\
Indian Institute of Technology Hyderabad,
Kandi, Sangareddy 502285, India }
\maketitle
\section*{Abstract}
Symmetry breaking is essential for biological microswimmers to achieve locomotion in viscous environments, enabling the generation of directional forces that sustain motion. 
Autophoretic microswimmers---such as isotropic active colloids and solubilizing droplets-- undergo spontaneous symmetry breaking of a chemical species distribution, which generates interfacial flows that drives persistent self-propulsion.
A major complexity in modeling such autophoretic microswimmers stems from the coupling between the evolution of the chemical concentration and the flow fields around the microswimmer through the non-linear advective transport of the chemical species.
Here, we present a novel numerical framework for simulating isotropic autophoretic microswimmers whose propulsion arises solely from self-generated chemical gradients, without any built-in geometric or chemical anisotropy.
The method employs a high-accuracy pseudospectral solver that resolves the coupled advection–diffusion–Navier-Stokes equations on a fixed Eulerian grid.
Slip velocities emerge self-consistently from instantaneous concentration gradients at the particle surface, and the resulting propulsion is coupled to the generated flow through a regularized stresslet representation of force- and torque-free microswimmers.
This approach simultaneously captures nonlinear advection, chemo-hydrodynamic feedback, and many-body interactions within a single computational framework.
Comparing with independently performed experiments using self-propelling active droplets, we demonstrate that the numerical framework reproduces complex motility patterns characteristic of autophoretic microswimmers, including disordered swimming at high viscosities and chemotactic avoidance during pairwise interactions.
The proposed framework reliably captures the underlying mechanisms governing emergent autophoretic dynamics.

\section{Introduction}
\label{sec:headings}

Over the years, synthetic microswimmers have emerged as autonomous agents for executing intricate tasks, like metal ion detection and collection \cite{ban2018motion}, targeted drug delivery \cite{kagan2010rapid, ma2022magnetosome}, assisted fertilization \cite{bunea2020recent}, and water remediation \cite{li2022self}.
The design and function of these micro/nano-machines often draw direct inspiration from microorganisms, which in turn have evolved sophisticated strategies to achieve efficient propulsion in complex environments \cite{ghosh2009controlled}. 
While engineered microswimmers can be actuated by external fields, such as electric \cite{han2018engineering, han2021low}, magnetic \cite{tierno2008magnetically,martin2023magnetic}, or optical stimuli \cite{bunea2021light}, the practical implementation of such methods faces significant hurdles stemming from miniaturization constraints, fabrication complexity, and diminished performance outside of controlled laboratory settings \cite{ebrahimi2021magnetic,joh2021materials}.
In contrast, chemically active microswimmers offer a self-sustained alternative by converting chemical free energy directly into autonomous motion without requiring any external actuation \cite{moran2017phoretic, maass2016swimming, michelin2023self, birrer2022we}. 
Their propulsion arises either from phoretic mechanisms, like self-diffusiophoresis or self-electrophoresis (as in Janus microswimmers) \cite{paxton2004catalytic, sundararajan2008catalytic, velegol2016origins, illien2017fuelled}, or from solubilization of one chemical species in another (as in active droplets) \cite{maass2016swimming,michelin2023self,toyota2009self, thutupalli2011swarming, peddireddy2012solubilization, herminghaus2014interfacial, izri2014self}, whereby interfacial chemical interactions create gradients in chemical concentration or surface tension which trigger interfacial flow.
Once the interfacial flow sets in, the particle self-propels in order to satisfy the non-inertial (zero net hydrodynamic force) constraints of the low Reynolds number regime. 
Note here that for a Janus-type microswimmer, a material (chemical) asymmetry must be externally engineered to create persistent chemical gradients that drive continuous self-propulsion \cite{paxton2004catalytic,howse2007self, zottl2016emergent, zhang2017janus, zhang2017self, katuri2017designing, ebbens2018catalytic,campbell2019experimental, liebchen2021_interactions, zottl2023_modeling}. 
The fact that the asymmetry necessary for self-propulsion needs to be engineered during synthesis of the microswimmer in the form of a catalytic metal deposited over part of its surface makes the fabrication of Janus microswimmers complex and infrastructure dependent. 
These are relatively difficult to fabricate in large numbers with uniform and controllable surface properties. 
Hence, the behaviour of Janus microswimmers is often less reproducible and less reliable in complex environments \cite{su2019janus}.
These limitations of Janus microswimmers have motivated a growing interest in chemically isotropic autophoretic microswimmers \cite{golestanian2007designing}.

Autophoretic microswimmers achieve self-propulsion through \textit{spontaneous symmetry breaking} of a chemical species concentration triggered by advective perturbations, without the need for any engineered chemical asymmetry \cite{michelin2013spontaneous}.
On one hand, for isotropic, chemically reactive colloidal particles, such spontaneous symmetry breaking of a chemical species created by surface chemical reactions leads to a phoretic slip velocity, which triggers interfacial flow and hence self-propulsion \cite{michelin2014phoretic, moran2017phoretic}.
On the other hand, for solubilizing droplets in a supramicellar surfactant solution, the spontaneous symmetry breaking of the filled micelle concentration leads to interfacial surfactant (interfacial tension) gradient, which drives interfacial flow triggering self-propulsion of the droplets \cite{toyota2009self, thutupalli2011swarming, peddireddy2012solubilization, izri2014self, suga2018self,herminghaus2014interfacial, maass2016swimming,michelin2023self, birrer2022we, kruger2016curling, moerman2017solute}. 
Both isotropic chemically reactive colloidal particles and active droplets are canonical examples of autophoretic microswimmers.
It is important to note here that for autophoretic microswimmers far away from any boundary, the absolute necessity of advection for the symmetry breaking of the chemical species couples the evolution of the chemical concentration and the flow fields around the microswimmer through the non-linear advection terms in the species conservation equation, besides the pertinent hydrodynamic boundary conditions.
This nonlinear chemo-hydrodynamic coupling gives rise to rich emergent dynamics and places autophoretic microswimmers within the broader class of active matter systems, where continuous local energy input and nonlinear interactions lead to complex nonequilibrium behavior. 

In unbounded domains, theoretical and experimental studies have shown that increasing advective transport relative to diffusive effects, typically quantified by the increasing Péclet number, can destabilize steady swimming of autophoretic microswimmers due to chemo-hydrodynamic interactions giving rise to unsteady and even chaotic trajectories \cite{hu2019chaotic, farutin2022reduced,  michelin2023self, hokmabad2021emergence, morozov2019nonlinear, dwivedi2021solute}.
In geometric confinements or during interaction between microswimmers, chemo-hydrodynamic coupling is further accentuated resulting in trapping of a microswimmer in the flow field of another microswimmer  or near obstacles \cite{jin2019fine, izzet2020tunable}.
Chemo-hydrodynamic interactions can also result in emergent collective behaviour, including formation of swimming chains \cite{thutupalli2018flow, kumar2024emergent}, dynamic clusters \cite{thutupalli2011swarming, krueger2016dimensionality, varma2018clustering}, transient caging phenomena \cite{hokmabad2022chemotactic}, rotating clusters and self-organized vortical structures \cite{hokmabad2022spontaneously, cantisan2023rotating}, and emergent active turbulence \cite{yang2024shaping}. 
Such interesting emergent phenomena stem from the nonlinear interaction between fluid flow and chemical signals produced by the autophoretic microswimmers. 
However, at the same time, non-linearity of the interaction makes the prediction of such emergent dynamics through theoretical and numerical modeling extremely challenging. 

The principal challenge in modeling autophoretic microswimmers lies in the fact that these models must resolve the coupled Stokes flow and solute advection–diffusion with moving boundaries, since surface concentration gradients directly generate the slip velocity that drives propulsion \cite{anderson1989, golestanian2007designing}. 
 For the case of a single autophoretic microswimmer near a wall, or for two-swimmer interactions, the full chemo-hydrodynamic coupling can be addressed in bi-spherical coordinates \cite{lippera2020collisions,PhysRevFluids1} using a combination of analytical and numerical methods. 
However, given the apparent complexity of this modeling technique, it will be difficult to extend the framework to capture emergent collective behaviour of autophoretic microswimmers.
To address the collective behaviour, often the microswimmers are idealized as point sources instead of being finite-sized \cite{lippera2021alignment}, or the framework is simplified by assuming rapid chemical diffusion and viscous momentum transport \cite{singh2019competing}. 
Although the simplified models enable the implementation of efficient large-scale solvers \cite{delmotte2024scalable}, it comes at the cost of failing to capture emergent phenomena driven by nonlinear chemohydrodynamic coupling. 

Moving-boundary problems in the Stokes regime are commonly solved using boundary integral methods (BIM) based on Green’s functions.
BIM is the standard tool in microhydrodynamics, and works very well for viscous flows around passive particles \cite{pozrikidis1992boundary}.
However, for autophoretic microswimmers, resolving the chemical field throughout the fluid is computationally costly and typically scales as $\mathcal{O}(N^2)$ with the number of particles.
Consequently, fully coupled chemo-hydrodynamic BIM approaches are hard to scale to large swimmer populations or strongly nonlinear regimes. Other numerical approaches have also been developed to handle fluid flow and solute transport with moving boundaries.
The embedded boundary or cut-cell method \cite{picella2022confined} locally adjusts a Cartesian grid around each particle, giving sharp boundary resolution while leaving the rest of the grid fixed \cite{mccorquodale2001cartesian, schneiders2016efficient}. 
The overlapping mesh method (OMM) divides the domain into multiple body-fitted, overlapping subdomains and interpolates values at the interfaces, allowing each subdomain to move independently \cite{chesshire1990composite, henshaw2008parallel, prewitt2000parallel, merrill2019moving}. 
Recent high-accuracy numerical frameworks combining overlapping mesh methods with boundary-integral solvers \cite{gou2025computational} can accurately resolve moving boundaries and surface flux conditions. Similarly, immersed boundary methods have been successfully used for large-scale simulations of interacting active particles \cite{yang2024shaping}. However, these methods often require repeated interpolation, iterative coupling steps, or explicit reconstruction of moving boundaries at every timestep, which can make long-time simulations and exploration of wide parameter ranges computationally expensive. Moreover, in such fully coupled formulations, the chemo-hydrodynamic interaction is often resolved through iterative preconditioned procedures whose convergence can depend sensitively on relaxation parameters and may require multiple inner iterations within each timestep. To overcome these challenges, simpler fixed-grid approaches like the immersed boundary method  \cite{cs2002immersed, wang2020immersed, yang2024shaping} are often preferred.
Here, boundary forces are represented as source terms smoothly spread over nearby points, which avoids explicit boundary tracking and allows stable simulations of many particles.
While this approach improves numerical stability and circumvents the need for explicit moving-boundary treatment of the flow field, the diffuse interface still requires the enforcement of surface flux and slip conditions for the chemical field at the particle boundary, which remains a non-trivial challenge. Alongside these grid-based methods, particle-based frameworks such as multiparticle collision dynamics \cite{noguchi2004fluid,noguchi2005shape}, dissipative particle dynamics \cite{hoogerbrugge1992simulating,fedosov2010multiscale,xu2023coarse} and smoothed particle hydrodynamics \cite{monaghan1994,tanaka2005microscopic,hosseini2009particle} provide meshfree formulations that naturally capture mesoscale flows and complex interface physics.

In the proposed framework, the incompressible Navier--Stokes equations are solved using a pseudospectral FFT-based method with $N\log N$ scaling, while the solute concentration evolves according to the advection--diffusion equation on the same Eulerian grid.
At each timestep, the swimmer’s chemical activity is modeled by a smoothly distributed source centered at the particle, allowing the solute field to be evolved on a fixed grid without imposing surface flux boundary conditions at any moving surface --- thereby circumventing the key limitation that persists even in immersed boundary approaches. The resulting concentration gradient is computed on the grid and interpolated to the particle surface to evaluate the phoretic slip velocity \cite{anderson1989,golestanian2007designing}.
The resulting swimming speed and swimming orientation then emerge naturally via the Lorentz reciprocal theorem \cite{michelin2014phoretic}, ensuring a fully self-consistent coupling between the chemical and hydrodynamic fields. Each particle is represented as a point stresslet whose strength depends on its instantaneous swimming velocity \cite{lauga2016stresslets}, which is regularized and spread onto the fluid grid as a forcing term \cite{gupta2018conditional}, allowing its hydrodynamic influence to be incorporated globally without explicitly resolving particle boundaries. 
The updated flow subsequently advects the solute field, and the coupled system is advanced explicitly in time without the need for inner convergence iterations or moving meshes. 
Importantly, neither the particle velocity nor its orientation is prescribed; starting from a minimal initial perturbation that breaks the isotropic symmetry of the concentration field, both the magnitude and direction of propulsion arise autonomously from the evolving solute gradients.
The present framework intentionally adopts a coarse-grained representation that does not explicitly enforce surface flux or local slip boundary conditions. Consequently, near-field details such as thin solute boundary layers \cite{michelin2014phoretic} and lubrication-scale interactions are not fully resolved. However, the framework retains the dominant nonlinear coupling between solute advection and hydrodynamic flow responsible for large-scale chemo-hydrodynamic interactions, instability-driven motion, and collective dynamics. By avoiding moving meshes and inner iterative solvers, the method enables efficient exploration of long-time dynamics and parameter regimes that are computationally challenging for fully resolved formulations.
To assess the efficacy of the proposed framework, we compare the numerically simulated dynamics of the autophoretic microswimmer with the swimming dynamics of autophoretic active droplets observed in independent experiments performed by us.
Furthermore, we also test to what extent this new framework can capture the underlying chemo-hydrodynamic interactions that lead to some established complex motility patterns for autophoretic microswimmers, such as emergent destabilized swimming with increasing viscosity of the swimming medium \cite{hokmabad2021emergence} and pair-wise interactions mediated by chemotactic avoidance \cite{hokmabad2022chemotactic}.
To this end, we also compare the predictions from the numerical simulations with independently performed experimental results. Through these comparisons, we demonstrate that the proposed framework captures the essential chemo-hydrodynamic mechanisms governing complex motility patterns in autophoretic microswimmers.

The paper has been organised as follows-- \S ~\ref{sec:numerics} describes the numerical simulations, \S ~\ref{sec:experiment} outlines the experimental methodology, \S ~\ref{sec:results} presents and discusses the results, and \S ~\ref{sec:conclusions} provides the concluding remarks.

\section{Numerical simulations}
\label{sec:numerics}

\subsection*{Model framework}
\label{sec:model}
We consider a chemically active circular disk of radius $a$, immersed in a Newtonian fluid of kinematic viscosity $\eta$, representing a two-dimensional analogue of an autophoretic microswimmer.
Hereafter, the disk is referred to as a `microswimmer'.
Although we solve the full Navier--Stokes equations, the dynamics of such microswimmers are dominated by viscous forces, and falls within the Stokesian regime (\S \ref{sec:fluid}).
The concentration of the solute being created or emitted by the microswimmer is initialized uniformly throughout the domain, with small random perturbations to eventually induce spontaneous symmetry breaking. 
The microswimmer emits solute in an isotropic manner, mimicking the formation of a chemical species by an autophoretic microswimmer, like the emission of filled micelles by an active droplet solubilizing in a surfactant solution having concentration greater than the critical micellar concentration (CMC) \cite{maass2016swimming,michelin2023self}.
However, this isotropic state becomes unstable in the presence of perturbations, leading to an asymmetric solute distribution, and consequently, tangential concentration gradients along the surface. 
This generates a slip velocity proportional to the surface concentration gradient, resulting in self‑propulsion while the swimmer remains force‑free and torque‑free (\S \ref{sec:concentration}–\ref{sec:conc_emission}). 
The self-propulsion of the microswimmer perturbs the surrounding fluid, which in turn redistributes the solute, thereby coupling solute transport and the hydrodynamic interactions (\S \ref{sec:stresslet}).
Thus, the microswimmer motion, the solute field, and the fluid flow are determined self-consistently through their mutual coupling, as described in the following subsections.

\subsection{Fluid flow and vorticity formulation}
\label{sec:fluid}
To model the flow, we solve the 2D incompressible Navier–Stokes equations in vorticity–streamfunction form:

\begin{align}
\partial_t \omega + \mathbf{u} \cdot \nabla \omega &= \nu \nabla^2 \omega + f^a_\omega, \label{eq:vorticity}\\
\nabla^2 \psi &= -\omega, \label{eq:poisson}
\end{align}

where $\mathbf{u} (\mathbf{x},t)= \nabla^\perp \psi = (-\partial_y \psi, \partial_x \psi)$ is the velocity field, $\omega = (\nabla \times \mathbf{u}) \cdot \hat{\mathbf{z}}$ is the vorticity and $\hat{\mathbf{z}}$ is the unit vector normal to the plane of flow, and $\nu$ is the kinematic viscosity. The term $f^a_\omega$ accounts for triggering a disturbance in the surrounding fluid due to the self-propulsion of the microswimmer (see \S. ~\ref{sec:stresslet} for details).

\subsection{Solute emission and concentration field evolution}
\label{sec:conc_emission}

\begin{figure*}
  \centering
  \includegraphics[width=\textwidth]{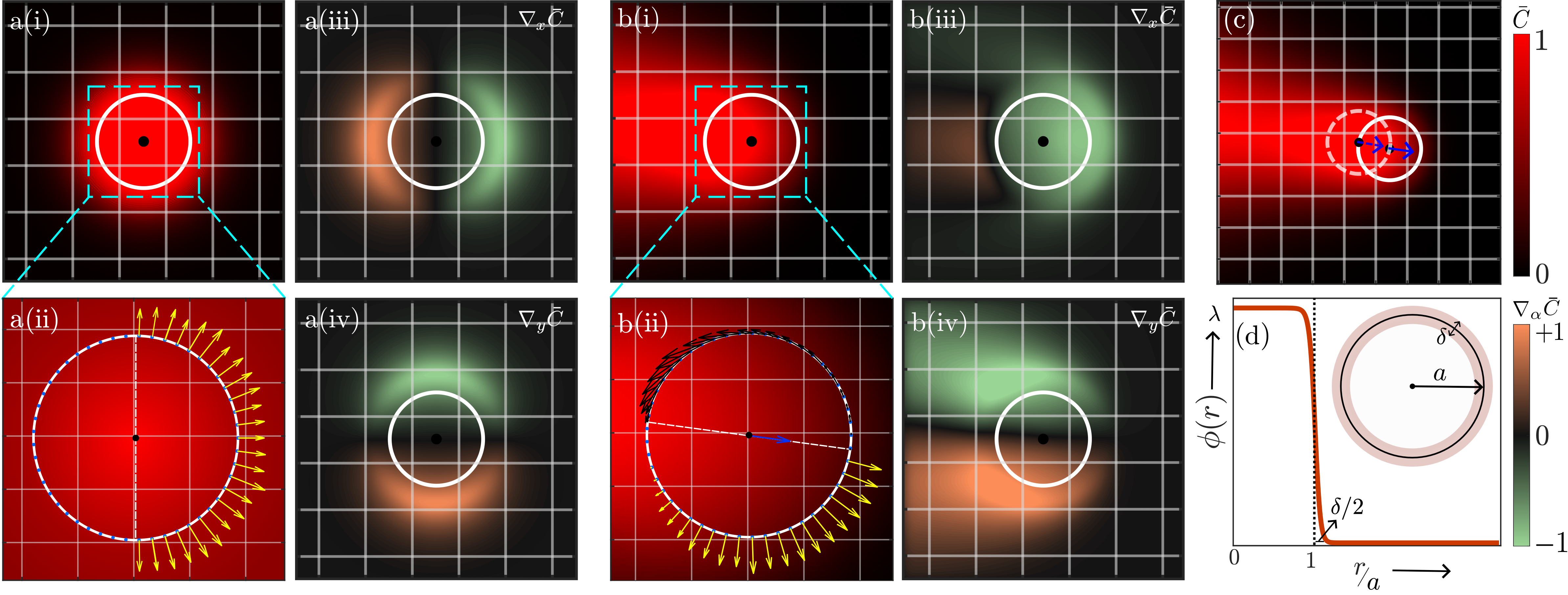} 
\caption{
\textbf{Schematic of the autophoretic microswimmer model.}
a(i) At low P\'eclet number ($\Pe$), diffusion dominates, producing an isotropic solute field $(\bar{C}= C/C_{max})$ with uniform gradients throughout the domain. The Cartesian components of the concentration gradient shown in panels a(iii)–a(iv) remain perfectly symmetric. In panel (a)(ii), the microswimmer is split along its diameter (indicated by the vertical, white dashed line) for visualization. The left half shows the slip velocity evaluated on the discretized surface (black arrows, almost invisible to the eye), which exhibits no tangential variation and therefore results in zero propulsion (i.e., the absence of surface gradients precludes any net phoretic driving). The right half shows the nearly uniform normal flux vectors (yellow arrows).
b(i) At higher $\Pe$, advection becomes significant and the solute distribution develops anisotropy. Panels b(iii)–b(iv) show the resulting asymmetric cartesian components of the gradient of $\bar{C}$ . In panel (b)(ii), the microswimmer is split along the swimming direction (indicated by the white dashed line) for visualization. The upper half shows the slip velocity evaluated on the discretized surface (black arrows), which now exhibits a finite tangential variation and a non-zero mean component, thereby driving propulsion opposite to the region of strongest slip (i.e., from high to low surface concentration). The lower half shows the normal flux, which is no longer uniform along the surface; it is nearly zero at the posterior and increases toward the anterior.
(c) A zoomed-out view highlights the swimmer’s displacement relative to its self-generated concentration field. 
(d) Radial solute distribution $\phi(r)$ (red curve) around a microswimmer, modeled using a hyperbolic-tangent profile with surface concentration $\lambda$, particle radius $a$, and decay length $\delta$ that sets the distance over which the concentration relaxes to the background value. The inset schematic shows the microswimmer (black circle of radius $a$) and the surrounding decay layer (beige annulus of thickness $\delta$).
\label{fig:1}
}
\end{figure*}

Each microswimmer of radius `$a$' acts as a localized chemical source, continuously emitting solute into the surrounding fluid, whose concentration field subsequently evolves through the combined effects of swimmer activity-driven advection and molecular diffusion.

At the start of the simulation, the solute concentration is initialized as a uniform background field with a small random perturbation,
\begin{equation}
C(\boldsymbol{r}, t_0) = C_0 + \xi(\boldsymbol{r}),
\end{equation}
where $C_0$ is a constant reference concentration and $\xi(\boldsymbol{r})$ is an uncorrelated, low-amplitude noise field introduced to seed spontaneous symmetry breaking.
Note that in the absence of any flow, diffusion rapidly smooths out these perturbations.

Chemical activity for the particle is implemented by placing a localized solute source at the instantaneous  center of the microswimmer $\mathbf{r}_i(t)$. The microswimmer emits solute at a prescribed rate $\lambda$, with a spatial distribution modeled by a smooth hyperbolic-tangent profile (Fig.~\ref{fig:1}d),
\begin{equation}
\phi(\mathbf{r}) = \frac{\lambda}{2} \left[1 - \tanh\!\left(\frac{|\mathbf{r}-\mathbf{r}_i| - a}{\delta}\right)\right].
\label{eq:activity}
\end{equation} 
where $\delta$ controls the thickness of the solute boundary layer surrounding the swimmer. 
This regularization avoids discontinuities at the surface of the particle while preserving a well-defined source region of size $a$.

The time evolution of the solute concentration is governed by an advection--reaction--diffusion equation,
\begin{equation}
\partial_t C + \mathbf{u} \cdot \nabla C = D \nabla^2 C + \phi(\boldsymbol{r}),
\label{eq:advection_diffusion}
\end{equation}
where $D$ is the solute diffusivity and $\mathbf{u}$ is the fluid velocity field obtained from \S\ref{sec:fluid}. At each time step, newly emitted solute is added to the concentration field through the source term $\phi$. 
The spatial distribution of the solute concentration field has been discussed in detail in Appendix Fig.~\ref{fig:supp2}.

The interplay of activity, advection, and diffusion generates strong concentration gradients in the immediate vicinity of the microswimmer, while farther away from the particle the solute distribution relaxes smoothly toward the background concentration (Appendix Fig.~\ref{fig:supp2}).
As shown in Fig.~\ref{fig:1}(a), diffusion dominates at low P\'eclet number ($\Pe$) (see \S ~\ref{sec:results} for the definition of $\Pe$), leading to an essentially isotropic concentration field.
At higher P\'eclet number (Fig.~\ref{fig:1}(b)), advection distorts the solute distribution and produces anisotropic gradients along the swimmer surface, which drive phoretic slip and self-propulsion (Fig.~\ref{fig:1}(c)) (§\ref{sec:concentration}).

\subsection{Phoretic slip and self-propulsion}
\label{sec:concentration}

Gradients in the solute concentration along the surface of the microswimmer generate a local phoretic slip velocity (\cite{anderson1989,michelin2013spontaneous})
\begin{equation}
\mathbf{u}_s = M \nabla_{\parallel} C ,
\label{eq:phoretic_slip}
\end{equation}
where $M$ is the phoretic mobility, $\nabla_{\parallel}$ is the surface gradient, and $C$ is the solute concentration evaluated at the microswimmer boundary. 
For an isotropic concentration field, the  surface-averaged slip vanishes and the swimmer remains stationary.

The instantaneous swimming velocity of a force-free microswimmer can be evaluated using the Lorentz reciprocal theorem \cite{stone1996propulsion}, and is determined solely by the surface slip velocity (Eq. \ref{eq:phoretic_slip}).
In two dimensions, this relation reduces to a line integral along the particle boundary $\partial B$,
\begin{equation}
    \mathbf{u}_0 = -\frac{1}{2\pi a} \int_{\partial B} \mathbf{u}_s \, dl
    = -\frac{M}{2\pi a} \int_{\partial B} \nabla_{\parallel} C \, dl.
    \label{eq:translation}
\end{equation}
The negative sign indicates that the swimmer moves opposite to the surface slip, corresponding to an anti-chemotactic response for $M>0$.

At low $\Pe$, diffusion suppresses concentration anisotropies, resulting in nearly uniform surface gradients and vanishing propulsion (Fig.~\ref{fig:1}(a)).
In this isotropic case, the normal flux of the concentration at the surface of the microswimmer remains constant, as shown in Fig.~\ref{fig:1}(a(ii)) (yellow arrows in the right half). 
As the $\Pe$ increases, advection distorts the solute field and generates anisotropic surface gradients, leading to a finite mean slip (black arrows in the top half in Fig.~\ref{fig:1}(b(ii))) and sustained self-propulsion (Fig.~\ref{fig:1}(c)). 
Advection also affects the normal flux of the solute at the surface of the microswimmer, rendering it non-uniform.
As shown in Fig.~\ref{fig:1}(b(ii)) (yellow arrows in the bottom half), the flux remains finite and comparable to the isotropic case in the anterior region, while it is suppressed in the posterior region. 
This behavior contrasts with many theoretical models of microswimmers, where a spatially uniform normal flux along the boundary of the microswimmer (Neumann boundary condition) is imposed \cite{michelin2013spontaneous, picella2022confined, hu2019chaotic, zhu2023self, yang2024shaping}. 
Such a theoretical idealization may hold at the scale of the interfacial layer thickness.
However, our observations correspond to length scales comparable to the characteristic length scale of the microswimmer.
At these resolved scales, advection associated with self-propulsion leads to accumulation of solute in the posterior region, which naturally reduces the local concentration gradient, and consequently, suppresses the outward diffusive flux. 
In contrast, the anterior region remains relatively depleted of the solute, sustaining an almost uniform outward flux.
A similar front–back asymmetry in solute distribution is illustrated in schematic representations of self-propelling active droplets \cite{hokmabad2021emergence} (see Fig.~1 therein).
Although the propulsion mechanism in such systems is Marangoni-driven, the schematic highlights the general role of advection in shaping the concentration field. 
This asymmetry in solute distribution naturally gives rise to a non-uniform normal flux, with reduced flux at the posterior and finite, almost constant, diffusive flux at the anterior.

To evaluate the surface integrals numerically, the circular boundary of the microswimmer is discretized into $n$ uniformly spaced points. At each point, the surface concentration gradient is obtained by interpolating the grid-based solute field, and the swimming velocity is approximated as

\begin{equation}
    \mathbf{u}_0 \approx -\frac{M}{n} \sum_{k=1}^{n} (\nabla_{\parallel} C)_k
    \label{eq:discrete_velocity}
\end{equation}

where $(\nabla_{\parallel} C)_k$ is the surface gradient at the $k$-th boundary point. 

Once a finite swimming velocity is established, the instantaneous position of the microswimmer center $\boldsymbol{r}_i$ is updated according to
\begin{equation}
\partial_t \boldsymbol{r}_i = \boldsymbol{v}_d
\end{equation}
where the total velocity $\boldsymbol{v}_d$ of the microswimmer is
\begin{equation}
\boldsymbol{v}_d = \mathbf{u}_0 + \mathbf{u}(\boldsymbol{r}_i) + \sum_{\substack{i \ne j}} \mathbf{F}_{ij}
\label{eq:position_update}
\end{equation}
Here, $\mathbf{u}(\boldsymbol{r}_i)$ represents the contribution to the microswimmer velocity by the ambient flow field (calculated at the centre of the microswimmer), and $\mathbf{F}_{ij}$ accounts for steric interactions with neighboring microswimmers. 

Steric repulsion is modeled as a short-ranged, linearly decaying force,
\begin{equation}
\boldsymbol{F}_{ij} =
\begin{cases}
F_0 \left( 1 - \dfrac{r_{ij}}{2a} \right) \mathbf{e}_{ij}, & r_{ij} < 2a, \\
0, & \text{otherwise},
\end{cases}
\label{eq:steric}
\end{equation}
where $F_0$ is the repulsion strength, $r_{ij} = |\boldsymbol{r}_j - \boldsymbol{r}_i|$ is the distance between the microswimmer centers, and $\mathbf{e}_{ij} = (\boldsymbol{r}_j - \boldsymbol{r}_i)/r_{ij}$ is the unit vector pointing from microswimmer $i$ to $j$. For an isolated swimmer in the absence of ambient flow, $\boldsymbol{v}_d=\mathbf{u}_0$.

\subsection{Hydrodynamic signature and stresslet modeling}
\label{sec:stresslet}

The self-propulsion of a microswimmer also disturbs the surrounding fluid.
Microswimmers swim without exerting a net force or torque on the fluid, and as a result the leading-order far-field flow has the symmetry of an axisymmetric force dipole, known as a stresslet \cite{batchelor1970stress, ishikawa2007hydrodynamic}. 
This stresslet can be calculated directly from the surface slip velocity using the reciprocal theorem, without solving for the full flow field \cite{lauga2016stresslets}.
Following this approach, we model the disturbance due to each microswimmer in our simulations as that of a point stresslet centered at its instantaneous position $\mathbf{r}_i$ (the exact strength of the stresslet is discussed in sub-section~\ref{sec:numerical})  
\begin{equation}
\sigma_0 \propto \alpha a^2 \nu u_0(t),  \label{eq:stresslet_0}
\end{equation} 
where $\alpha>0$ for pullers and $\alpha<0$ for pushers. 
However, since a point stresslet corresponds to a delta-function stress distribution, we subsequently distribute it smoothly into the surrounding fluid using a Gaussian kernel \cite{maxey2001localized,lomholt2003force,lushi2013modeling, lushi2014fluid,gupta2018conditional}. 
This regularization ensures smooth coupling between the microswimmer and the fluid (in a way addressing the finite size of the microswimmer), improves numerical stability, and preserves the essential character of the stresslet while remaining computationally efficient.

The resulting regularized stress field is written as
\begin{equation}
\label{eq:stress_gaussian}
\mathbf{S}(\mathbf{r},t) =
\sum_{i=1}^{N_p}
\frac{\sigma_0}{(2 \pi \epsilon^2)^{3/2}}
 \left( \hat{\mathbf{e}}_i \hat{\mathbf{e}}_i - \tfrac{1}{2}\mathbf{I} \right)
\exp\!\left[-\tfrac{|\mathbf{r}-\mathbf{r}_i|^2}{2\epsilon^2}\right],
\end{equation}
where $\hat{\mathbf{e}}_i$ (Fig.~\ref{fig:2}(b),inset) is the swimmer orientation and $\epsilon$ the Gaussian width, effectively representing the size of the swimmer. ${N_p}$ is the total number of microswimmers.
Although a swimmer does not act directly on itself through this stress, the flow field generated by the regularized stress advects the surrounding solute, thereby coupling the chemical and hydrodynamic fields. In the simulations, the active forcing that drives the fluid motion is obtained from the divergence of the regularized stress field,
\begin{equation}
\mathbf{f}^a = \nabla \cdot \mathbf{S},
\end{equation}
which, in the streamfunction--vorticity formulation, enters as  
\begin{equation}
f_\omega^a = \nabla \times (\nabla \cdot \mathbf{S}).
\end{equation}

This formulation enables a self-consistent coupling between swimmer activity, hydrodynamic interactions, and solute transport while remaining computationally tractable for large-scale simulations.

\subsection{Numerical methodology}
\label{sec:numerical}   

We perform direct numerical simulations (DNSs) of Eqs.~\eqref{eq:vorticity}–\eqref{eq:advection_diffusion} in a doubly periodic square domain $[0,8\pi]^2$ using a pseudo-spectral method \cite{canuto2006spectral,gupta2014elliptical}. We represent the flow and concentration fields on a uniform $1024^2$ grid. We compute spatial derivatives in Fourier space and evaluate nonlinear terms in physical space via transform–multiply–inverse-transform. We remove aliasing using the two-thirds truncation \cite{orszag1971elimination}, and FFTW provides fast transforms \cite{frigo2005design}. Grid-refinement tests at $N=512,1024,2048$ show negligible changes in statistics and swimmer velocities, so we adopt $N=1024$.

We solve the two-dimensional incompressible Navier–Stokes equations using a vorticity–streamfunction formulation \cite{perlekar2011persistence,gupta2014elliptical,pandit2017overview}, with $\mathbf{u}=(\partial_y\psi,-\partial_x\psi)$ ensuring incompressibility. We advance time using a second-order exponential time-differencing Runge--Kutta (ETD-RK2) \cite{press2007numerical}, which treats diffusive terms exactly and nonlinear terms explicitly. We use a fixed step $\Delta t=10^{-3}$ that satisfies the Courant–Friedrichs–Lewy (CFL) condition $\max|\mathbf{u}|\Delta t/\Delta x\lesssim1$, where $\Delta x=8\pi/N$.

We vary the viscosity over $\nu\in[8\times10^{-3},5]$ and prescribe the diffusivity as $D=D_0/\nu$, ($D_0=8\times10^{-3}$), which tunes the $\Pe$ number without destabilising the solver. The advection–diffusion implementation matches the four-roll mill benchmark \cite{thomases2009transition,solomon2003uniform}.

We obtain concentration gradients at the swimmer surface by bilinear interpolation from the Eulerian concentration gradient field to Lagrangian boundary points~\cite{gupta2014elliptical}.
These gradients are then used to compute the phoretic slip velocity along the surface (Eq.~\ref{eq:phoretic_slip}). 
Using the Lorentz reciprocal theorem, we subsequently determine the swimming velocity $\mathbf{u}_0$ (Eq.~\ref{eq:discrete_velocity}) of the microswimmer. 
The computed velocity is further used to estimate the stresslet strength through the relation $\sigma_0 = 8 \pi \alpha a^2 \nu u_0(t)$.
The proportionality constant of Eq.~\ref{eq:stresslet_0} is determined based on numerical experiments, where an order-of-magnitude agreement is obtained between the terminal swimming speed of the microswimmer and the induced flow disturbance in its neighbourhood.
In the present work, we take $\alpha = -1$ to mimic the pusher nature of the swimmer.
The prefactor $8\pi$ is of the same order as has been taken by Lauga et.al.~\cite{lauga2016stresslets}.
We introduce swimmer forcing as a regularised body-force density with Gaussian width $\epsilon=a/(6\sqrt{\pi})^{1/3}$ (\S.~\ref{sec:stresslet}), which provides smooth coupling on the grid~\cite{maxey2001localized}.
The solute activity layer (Eq.~\eqref{eq:activity}) has thickness $\delta\simeq5\Delta x$, which is well resolved at the chosen grid spacing.
The robustness of this choice, and its effect on the swimming dynamics, are discussed in Appendix (see Fig.~\ref{figure:supp1}).

The governing equations are solved in dimensional form. This preserves the direct physical meaning of parameters and simplifies coupling between solute transport, hydrodynamics, and swimmer activity when multiple length and time scales coexist. Dimensionless groups such as the Reynolds and $\Pe$ numbers are evaluated a posteriori to classify regimes.

We impose swimmer forcing as a body force rather than a boundary condition, which avoids explicit surface meshing and remains compatible with the spectral solver and periodic geometry, while reproducing the correct far-field stresslet structure \cite{lushi2013modeling,lushi2014fluid,gupta2018conditional}.

\section{Experimental Methodology}
\label{sec:experiment}
We compare the motility patterns for autophoretic microswimmers predicted by the numerical framework with results from independently performed experiments using self-propelling active droplets- a canonical example of an autophoretic microswimmer \cite{maass2016swimming,michelin2023self,toyota2009self, thutupalli2011swarming, peddireddy2012solubilization, herminghaus2014interfacial, izri2014self}.
Active droplets are droplets undergoing solubilization in an immiscible surfactant solution having concentration greater than the critical micellar concentration (CMC).
Spontaneous symmetry breaking of the surfactant concentration field, along the droplet's interface, during solubilization drives interfacial flows that propel the droplets.
As these droplets self-propel, they leave behind (emit) a chemical trail of a solute (filled micelles).
The emission of this solute trail leads to interesting emergent motility patterns for active droplets through chemo-hydrodynamic feedback, such as destabilized swimming with increasing viscosity of the swimming medium \cite{hokmabad2021emergence, dwivedi2021solute}, and pair-wise interaction mediated by chemotactic avoidance of the trail \cite{hokmabad2022chemotactic}. 
We also recreate these two complex motility patterns for active droplets in order to compare with the predictions from our numerical framework. 
Two things are to be noted here: (i) the phoretic slip boundary condition considered in the model is a simplification of the more complex combination of phoretic and Marangoni effects applicable for active droplet boundaries, as has been also done in the literature \cite{PhysRevFluids1, PhysRevFluids2}; (ii) we perform the experiments in quasi-2D setups, i.e. the active droplets are restricted to swim only along a plane and cannot exhibit 3D swimming. 
The last point makes the experimental observations compatible with the 2D simulation results.

\subsection{Generation of active droplets}
We generate the active (self propelling) oil droplets using a microfluidic flow focusing device (\cite{guchhait2025flow}).
In the microfluidic flow focusing device, the two streams of 0.1 wt$\%$ aqueous surfactant (Trimethyl(tetradecyl)ammonium bromide, TTAB ; Sigma-Aldrich) solution pinch off the CB-15 liquid crystal oil (PureSynth) stream creating droplets of $2a \approx43.3 \pm 1.88 \ \mu$m in diameter. 
We generate and store the CB15 oil droplets in 0.1 wt$\%$ aqueous TTAB solution which is lower than the critical miceller concentration for TTAB (CMC=0.13 wt$\%$). 
During the experiments, we mix these oil droplets with 7.5 wt$\%$ aqueous TTAB solution to make them active, and subsequently introduce them into the reservoir.

\subsection{Fabrication of quasi-2D reservoir}
The reservoir is fabricated using standard photolithography and softlithography processes with dimensions of 5$\times$5 mm$^2$ and a height $(h)$ of 50$\mu$m. 
The height of the reservoir is designed to be such that $2a/h \sim 0.9$, so that the droplet is confined along the $Z$ direction and can only swim in the horizontal $X-Y$ plane, effectively resulting in a quasi-two-dimensional (quasi-2D) system. 
We use polydimethylsiloxane (PDMS) (SYLGARD 184, Dow Chemicals) mixed in the base polymer to the crosslinker ratio of $10:1$. 
We pour the PDMS prepolymer on the reservoir mold, and cure it in a hot air oven for 3 hours at 75 $^\circ$C. Similarly, we coat a cover-slip (160 $\mu$m in thickness, Corning) with the same prepolymer film and cure it for 3 hours at 75 $^\circ$C. 
Upon curing, we peel off the cured PDMS stamp, having the negative replica of the mold, and create the inlet and outlet reservoirs using sharp punctures. 
We then bond the PDMS stamp to the glass slide coated with the PDMS film using a plasma oxidation process. 

\subsection{Bright field and fluorescence microscopy}
The trajectories of the active droplet are visualized using bright-field imaging using an inverted Nikon ECLIPSE Ti microscope. 
We track the active droplet using a CMOS monochrome digital camera integrated with the microscope (IDS Imaging), having an image resolution of $1920 \times 1200$ pixels. 
We record the image sequence at 25 frames per second, with a 20X objective having a spatial resolution of 0.2802 $\mu$m/pixel. 
The centroid of the active droplet is evaluated using an in-house MATLAB post-processing code. 
The instantaneous swimming velocity of the active droplet is calculated using the central difference method as $v_{d_x}(t) = (x(t+\Delta t) - x(t-\Delta t))/2\Delta t$ and similarly for $v_{d_y}$. 
Using $v_{d_x}$ and $v_{d_y}$, we calculate the instantaneous swimming orientation of the active droplet as $\psi= \tan ^{-1}(v_{d_y}/v_{d_x})$. 

We visualize the flow field generated by the active droplet using a high-resolution micro-Particle Image Velocimetry ($\mu PIV$) technique. 
The continuous medium (7.5 wt$\%$ aqueous TTAB solution) is seeded with fluorescent tracer particles of 500 nm in size (carboxylate-modified, Thermo Fisher Scientific) having excitation/emission wavelengths of 505/515 nm. The tracer particles are mixed in a ratio of $1:20$ by volume in the aqueous TTAB solution, and we sonicate the mixture to prevent coagulation of the tracer particles. 
We then add the generated CB15 oil droplets in this solution (containing 7.5 wt$\%$ aqueous TTAB solution mixed with tracer particles) in a ratio of $1:5$ by volume to make them active. 
To visualize and quantify the flow field generated by the active droplets, the tracer particles are excited using a $100$ W mercury lamp (Nikon, Model$:$ LH-M100C-1) attached to the inverted Nikon ECLIPSE Ti microscope. The required excitation wavelength is passed through a filter cube having filters compatible with the two peak excitation wavelengths of $487/562$ nm and emission wavelengths of $523/630$ nm (Chroma Technology, Model$:$ 59009 - ET - FITC/CY3). 
We record the image sequence at 25 fps, and post-process the image sequence in an open source platform integrated with MATLAB 2024a (\cite{stamhuis2014pivlab}). 

\subsection{Evaluation of chemical (filled micelle) kymographs}
To visualize the chemical (filled micelle) trail generated by the active droplets, we tag the CB15 oil with fluorescent Nile Red dye (Invitrogen™; excitation/emission: $550/640$ nm), and generate the droplets using identical microfluidic flow focusing device and store them in an aqueous surfactant solution having the surfactant concentration lower than the CMC level (0.1 wt$\%$).
As described previously, we make the droplets active by mixing them in the 7.5 wt$\%$ aqueous TTAB solution, and then introduce them into the quasi-2D reservoir. 
During the solubilization process of the active droplets, the empty micelles in the supramicellar surfactant solution take up oil and surfactant monomers from the interfacial region of the droplets and transform into filled micelles containing the dyed CB15 oil.
These dyed filled micelle trail emitted by the active droplet, as it self-propels,  can be visualized using the inverted microscope equipped with a dual-band filter (Chroma Technology, Model $:$ 59009 - ET - FITC/CY3).  
The image sequence is captured at 25 frames per second, and the exposure is set so that the image does not oversaturate (Fig.~\ref{fig:3}(d)).   

The image sequence is post-processed using an in-house Matlab 2024a code. 
We extract the pixel intensities $I(\theta)$ around the droplet microswimmer by traversing in a CCW direction $(\theta \in [0, 2 \pi])$ at a distance of 22$\mu m$ away from the active droplet interface (to avoid any noise due to intensity oversaturation).
The kymograph is generated by plotting the temporal variation of $I(\theta)$ around the active droplet over time (Fig.~\ref{fig:3}(e)).
Additionally, the diffusion profile of the filled micelles over time, in the chemical trail left behind as the active droplet swims away, is analyzed by extracting the pixel intensities ($I$) along a fixed line perpendicular to the local swimming orientation of the active droplet (Fig.~\ref{fig:3}(f)).
 
\subsection{Variation of the viscosity of the swimming medium}
The viscosity of the swimming medium (7.5 wt$\%$ aqueous TTAB solution) is varied by mixing glycerol (99$\%$; Sigma-Aldrich) at concentrations of 40$\%$ and 60$\%$ by volume in the total solution.
We characterize the viscosity of the swimming medium (aqueous surfactant solution) mixed with glycerol using a Anton Paar rheometer rotational rheometer (RheolabQC). 
The dynamic viscosity of aqueous surfactant solution mixed with 40$\%$ and 60$\%$ glycerol by volume are measured in the shear rate range of 10 $<$ $\dot\gamma$ $<$ 1500 s$^{-1}$, and are calculated to be 3.5 mPas and 11.5 mPas, respectively. 

\section{Results and discussions}
\label{sec:results}

To enable direct comparison between simulations and experiments, we nondimensionalize all results using equivalent microswimmer scales: spatial coordinates by the swimmer radius $a$ (simulation: $a=1$; experiments: $a=21.67\pm0.94~\mu$m), velocities by the characteristic swimming speed $U_0 \equiv (\langle v_d \rangle)$, i.e., average swimming speed measured in a low-viscosity medium with minimal fluctuations (simulation: $U_0\approx0.58$; experiments: $U_0=17.79\pm2.07~\mu$m/s in aqueous TTAB), and time by the advective scale $t_0=a/U_0$. Under this nondimensionalization, the P\'eclet number of the autophoretic swimmer is defined as $\mathrm{Pe} \approx U_0a/D(\nu)$, where $D(\nu)$ is the viscosity-dependent solute diffusivity.

In the simulations, spontaneous symmetry breaking of the concentration field over the swimmer length scale—and the resulting onset of self-propulsion—occurs once $\mathrm{Pe} \approx 1$. We denote this value as $\mathrm{Pe}_{c1}$, defined as the minimum Péclet number at which the microswimmer transitions from a stationary state to sustained self-propulsion. In three dimensions, $\mathrm{Pe}_{c1} = 4$ and is independent of the system size. In contrast, in two dimensions, $\mathrm{Pe}_{c1}$ depends on the box size due to the logarithmic decay of the concentration field, which makes the solution sensitive to the domain size~\cite{hu2019chaotic}. In our simulations, we observe that increasing the box size leads to a decrease in $\mathrm{Pe}_{c1}$, consistent with previous numerical results (see Appendix, Fig.~\ref{fig:supp3}). In the quasi-2D experimental system using a pure aqueous TTAB swimming medium, the corresponding onset occurs at $\mathrm{Pe}\approx4$. While this threshold is higher in experiments, the numerical prediction that self-propulsion emerges at $\mathrm{Pe}=\mathcal{O}(1)$ is consistent with the theoretical criterion for symmetry breaking in isotropic autophoretic microswimmers \cite{hu2019chaotic}.

\subsection{Predictions for swimming characteristics and hydrodynamic signature at low Péclet number}
\begin{figure} 
    \centering
    \centerline{\includegraphics[width=1.0\linewidth]{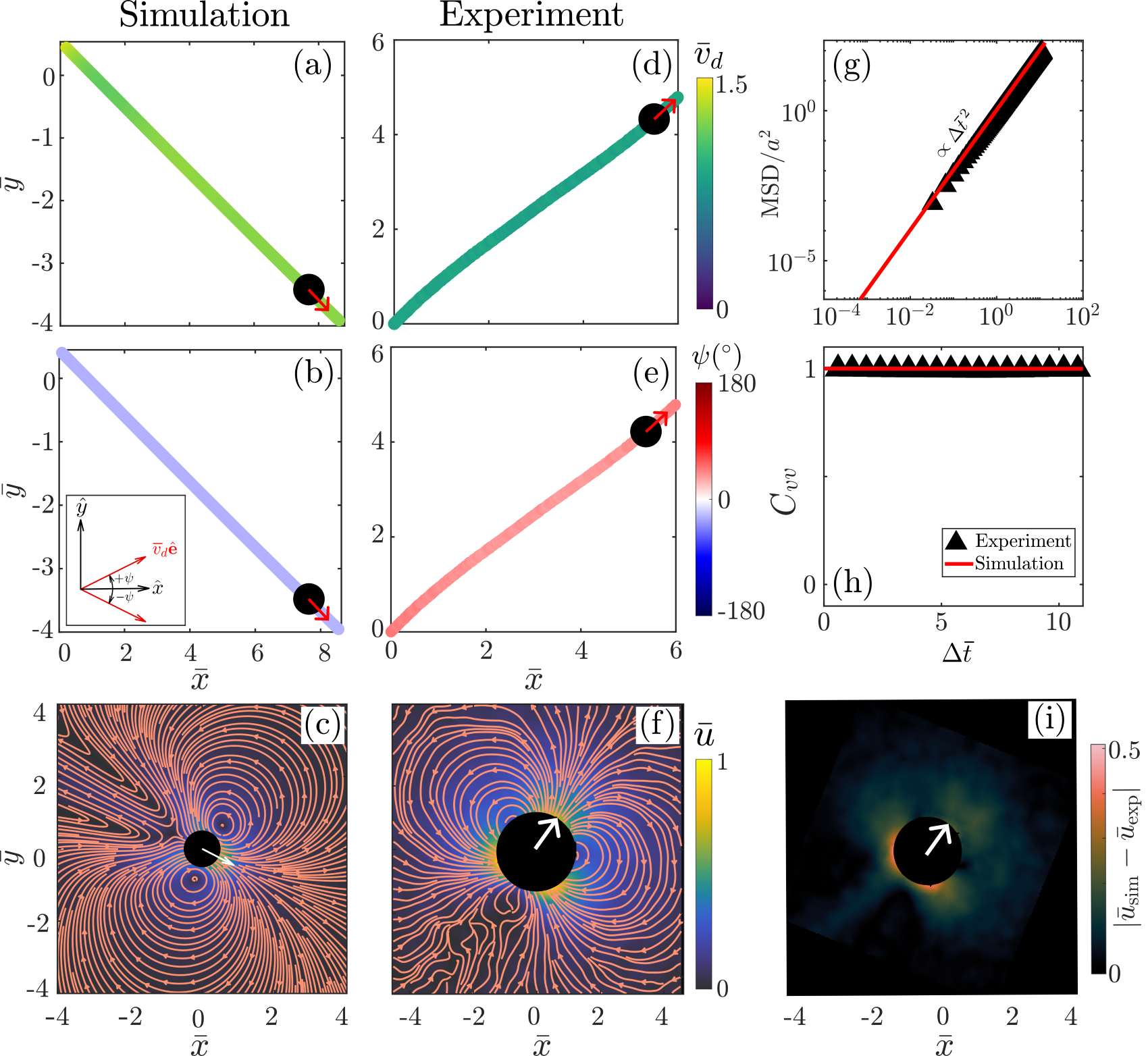}}
\caption{\textbf{Comparison between simulation and experimental results for microswimmer dynamics and for hydrodynamic signature at low P\'eclet number $(\mathrm{\bf Pe})$.} \textit{First column (Simulation):} Trajectory of the microswimmer color-coded (a) with the non-dimensional swimming speed $\bar{v}_{\mathrm{d}}$, and (b) with the swimming orientation as given by the orientation angle $\psi$ of the swimming velocity vector (inset). (c) The flow field generated by the microswimmer represented using streamlines and filled contour plot for the local velocity magnitude $\bar{u}$. The results shown here are for simulations at $Pe \approx 4$.  
\textit{Second column (Experiment):} (d)-(f) Corresponding experimental observations for $Pe = 4$.  
Temporal variations of the (g) mean-squared displacement (MSD) and (h) angular autocorrelation function $(C_{vv})$ for the swimming trajectory as obtained from the simulations (red line) and experiments (symbols). (i) The contour plot of the magnitude of the difference between simulation and experimental values of the local flow speed around the microswimmer swimming along identical direction.}
\label{fig:2}
\end{figure}

In simulations for $\mathrm{Pe} \approx 4$, the microswimmer swims along a persistent trajectory with non-dimensional steady speed $(\bar{v}_d)$ (Fig.~\ref{fig:2}(a)), and without alterations in the swimming orientation $\psi$ (Fig.~\ref{fig:2}(b); also see \href{https://drive.google.com/file/d/1V1NGNPW3XxptnuKSRVPj3R-g8Z1lzTTu/view?usp=drive_link}{movie S1} for $Pe \approx 6$, which exhibits dynamics equivalent to those observed at $Pe\approx 4$).
Here, $\psi$ is the orientation of the instantaneous swimming velocity vector $\boldsymbol{v_d}=v_d \boldsymbol{\hat{e}}=v_d (\cos \psi \; \boldsymbol{\hat{i}}+\sin \psi \; \boldsymbol{\hat{j}})$ (inset in Fig.~\ref{fig:2}(b)).
These swimming characteristics are consistent with the steady, unidirectional swimming exhibited by a model autophoretic microswimmer, such as an active droplet, in experiments for low $\mathrm{Pe}$ $(\approx4)$ (Figs.~\ref{fig:2}(d) and (e)).
Furthermore, we perform a quantitative comparison between the numerically simulated and experimentally observed swimming dynamics at low $\mathrm{Pe}$ using the mean–squared displacement (MSD) (Fig.~\ref{fig:2}(g)) and the angular autocorrelation function $C_{vv}$ (Fig.~\ref{fig:2}(h)). 
The MSD is computed as
\begin{equation}
\langle \Delta r^2(\Delta t) \rangle =
\big\langle \big[ \boldsymbol{r}(t_0+\Delta t)-\boldsymbol{r}(t_0) \big]^2 \big\rangle_{t_0},
\end{equation}
where the average is taken over all reference times $t_0$ and particle trajectories. 
The MSD for both the simulations (solid line) and the experiments (symbols) exhibit ballistic scaling $\langle \Delta r^2(\Delta t) \rangle \propto \Delta t^2$ (Figs.~\ref{fig:2}(g)), which is commensurate with the persistent propulsion over time observed in both the cases. 
The angular autocorrelation function is defined as
\begin{equation}
C_{vv}(\Delta t) =
\left\langle
\frac{\boldsymbol{v}_d(t_0+\Delta t)\cdot\boldsymbol{v}_d(t_0)}
     {\lvert\boldsymbol{v}_d(t_0+\Delta t)\rvert \,\lvert\boldsymbol{v}_d(t_0)\rvert}
\right\rangle_{t_0},
\label{eq:vacf}
\end{equation}
Here also, $C_{vv}$ is averaged over all reference times $t_0$ and trajectories. 
The auto-correlation function $C_{vv}$ computed from both simulation and experimental data remains constant about 1 (Fig.~\ref{fig:2}(h)), as expected for the unidirectional swimming.    

The numerical model predicts a dipolar flow field generated by the autophoretic microswimmer (Fig.~\ref{fig:2}(c)) during self-propulsion at low $\mathrm{Pe}$, as is also seen in the experiments with the active droplet (Fig.~\ref{fig:2}(f)).
To further pin-point the quantitative accuracy of the numerical prediction for the hydrodynamic signature, we show in Fig.~\ref{fig:2}(i) the contour plot for the magnitude of the difference (residual)  between the local values of the flow speed around a microswimmer, oriented in an identical manner, obtained from the simulations and experiments.

\subsection{Spatio-temporal evolution of the chemical trail emitted by the microswimmer}
\begin{figure}
    \centerline{\includegraphics[width=1.0\linewidth]{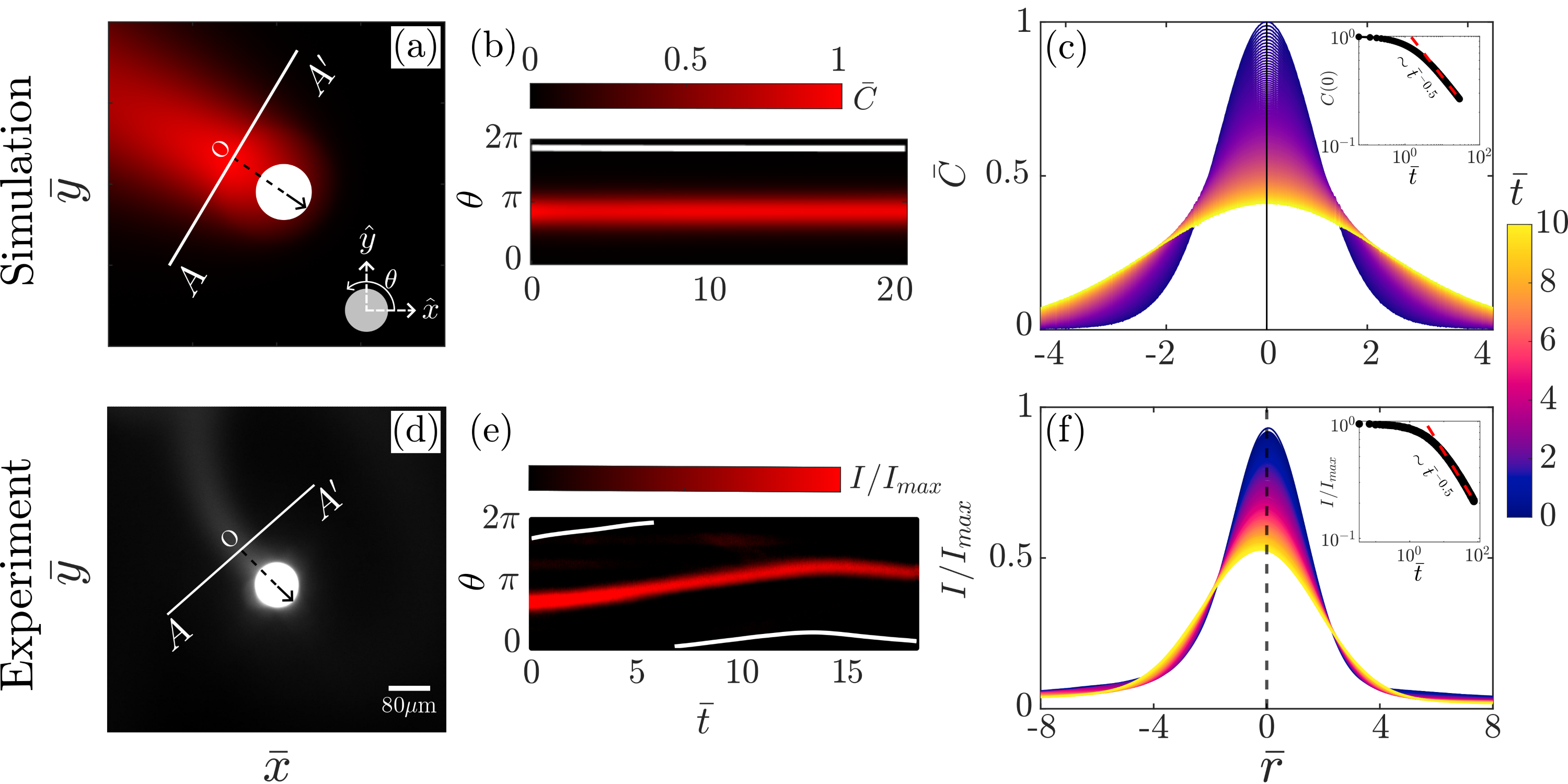}}
\caption{\textbf{Comparison between simulation and experimental results for the spatio-temporal evolution of the chemical trail ejected by the autophoretic microswimmer at low $\mathrm{\bf Pe} \approx 4$.}
\textit{Top row (Simulation):} (a) A simulation snapshot showing the self-generated chemical trail left behind by the microswimmer during its self-propulsion.   
(b) Kymograph of normalized circumferential concentration $\bar{C}(\theta,t)$; $\theta$ is measured counterclockwise from $+x$ (inset in a). The white line indicates the swimming orientation.
(c) Temporal evolution of $\bar{C}$ along a fixed line $\mathrm{AA^\prime}$ drawn orthogonal to the swimming direction (shown in a). 
Inset: log–log decay of concentration at a fixed point O on $\mathrm{AA^\prime}$, exhibiting $\propto t^{-1/2}$ scaling.
\textit{Bottom row (Experiment):} (d) Fluorescence micrograph showing the chemical (filled micelle) trail left by the active droplet.
(e) Corresponding kymograph of fluorescence intensity $I(\theta)$ (filled micelle concentration) around the active droplet periphery during its self-propulsion.
(f) Temporal evolution of the filled micelle concentration along a fixed line $\mathrm{AA^\prime}$ as in (d). The inset also shows the $\propto t^{-1/2}$ decay of the filled micelle concentration.}
\label{fig:3}
\end{figure}

The numerical framework captures the spatio-temporal evolution of the chemical trail generated by the autophoretic microswimmer (Fig.~\ref{fig:3}(a); \href{https://drive.google.com/file/d/1V1NGNPW3XxptnuKSRVPj3R-g8Z1lzTTu/view?usp=drive_link}{movie S1}), analogous to the oil-filled micelle trail observed in experiments (Fig.~\ref{fig:3}(d)).

The kymograph derived from the simulations (Fig.~\ref{fig:3}(b)) shows the evolution of the normalized circumferential concentration ($\bar{C}(\theta,t) = C/C_{max}$) during steady self-propulsion at low $\mathrm{Pe}\approx4$, where $\theta$ is measured counterclockwise from $+\hat{x}$ (inset of Fig.~\ref{fig:3}(a)). At low $Pe\approx4$, the swimmer maintains a persistent swimming direction, and the chemical wake remains uniform in thickness and diametrically anchored at the posterior apex, i.e. directly opposite to the instantaneous swimming orientation (white line).
The experimental kymograph (Fig.~\ref{fig:3}(e)) exhibits the same angular structure of the filled-micelle concentration, with the maximum consistently located diametrically opposite to the swimming orientation.

To quantify the evolution of the trail in the surrounding medium, we evaluate $\bar{C}(r,t)$ along a fixed line $AA^\prime$ drawn perpendicular to the swimming direction (Fig.~\ref{fig:3}(a)). The concentration profile follows a one-dimensional diffusive Gaussian (Fig.~\ref{fig:3}(c)):

\begin{equation}
C(r,t) \sim \frac{1}{\sqrt{4\pi D(\nu) t}} \exp\left( -\frac{r^2}{4D(\nu)t} \right),
\end{equation}
where $r$ is the distance from the center `O' of the trail along $AA^{\prime}$.
Furthermore, the inset in Fig.~\ref{fig:3}(c) shows that the peak concentration along $AA^{\prime}$ follows $\propto t^{-1/2}$ decay. Therefore, the evolution of the self-generated chemical trail in the swimming medium is diffusive in nature.
This is commensurate with the results from the experimentally observed evolution of the filled micelle trail left behind by the active droplet (Fig.~\ref{fig:3}(f)).
Note then that the time scale for the diffusion-driven evolution of the chemical trail in the swimming medium is $t_{diff} \sim a^2/D(\nu)\sim 8$, while the advective time scale for the propulsion of the microswimmer is $t_{adv} \sim a/U_0\sim 2$.
Since $t_{diff}>t_{adv}$, in our simulations, the medium is capable of retaining a memory of the microswimmer's trajectory as has been argued for experimental autophoretic systems \cite{hokmabad2022chemotactic}.

The agreement between numerical simulation and experimental results across microswimmer trajectories, flow fields, statistical measures, and the evolution of the self-generated chemical trail highlights the fact that the proposed numerical framework accurately captures the swimming dynamics of an autophoretic microswimmer, such as an active droplet.
The aforementioned discussions establish the foundation for investigating the efficacy of the proposed model in predicting more complex chemo-hydrodynamic interactions for autophoretic microswimmers. 

\subsection{Emergent complex motility at higher Péclet numbers}
\begin{figure}
   \centerline{\includegraphics[width=1.0\linewidth]{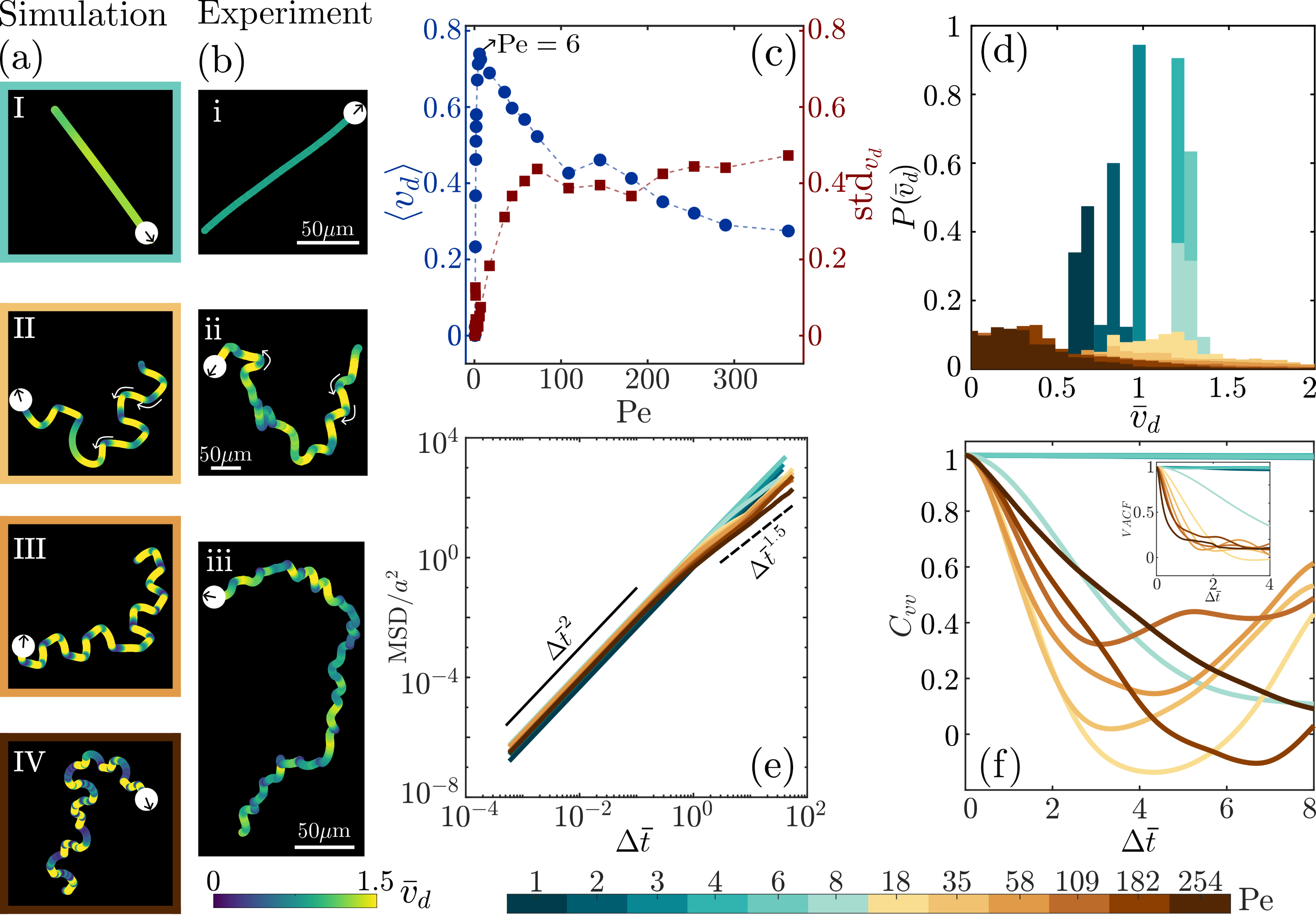}}
\caption{
\textbf{Emergent motility of the microswimmer with increasing viscosity of the swimming medium culminating in increasing $\mathrm{\bf{Pe}}$.}  
(a)(I-IV) Simulated microswimmer trajectories for increasing $\mathrm{Pe} \in \{6, 35, 58, 254\}$ color-coded by the non-dimensional propulsion speed $\bar{v}_d$.  
(b)(i-iii) Experimental trajectories for increasing $\mathrm{Pe} \in \{4, 15, 50\}$ corresponding to increasing viscosity of the swimming medium (achieved by adding increasing weight percentage of glycerol to aqueous TTAB solution- 0\%, 40\%, and 60\%). The white disks representing the active droplet are not to scale.
\textit{Statistical analyses of numerical simulation results:} (c) Variations of the average swimming speed$(\left<v_d\right>)$ and its standard deviation $(std_{v_d})$ with increasing $\mathrm{Pe}$.
(d) Probability distributions of the non-dimensional, instantaneous propulsion speed $(\bar{v}_d)$ for different $\mathrm{Pe}$.
(e) Variations of mean squared displacement (MSD), non-dimensionalized by $a^2$, with time interval $\Delta \bar{t}$ for increasing $\mathrm{Pe}$. 
(f) Variations of the angular autocorrelation function ($C_{vv}$) with time interval $\Delta \bar{t}$ for increasing $\mathrm{Pe}$. The inset shows the corresponding velocity autocorrelation function $(VACF)$, highlighting a monotonic decay of velocity correlations with time interval $\Delta \bar{t}$. }

\label{fig:4}
\end{figure}

Autophoretic microswimmers, like active droplets, adapt to increasing viscosity of the swimming medium, culminating in increasing $\mathrm{Pe}$, by spontaneously altering their motility from quasi-ballistic to chaotic (disordered) \cite{hokmabad2021emergence, dwivedi2021solute}.
We have performed experiments that recreate this transition in swimming motility of active droplets with increasing viscosity of the swimming medium achieved by adding increasing weight percentage of glycerol to the aqueous TTAB solution (Fig.~\ref{fig:4}(b)(i-iii)).
Note that the increasing viscosity of the swimming medium results in gradual increase in $\mathrm{Pe}$ due to the gradual decrease in $D(\nu)$.
Interestingly, the numerical simulation framework presented here predicts the emergence of such destabilized motion of an autophoretic microswimmer with increasing $\mathrm{Pe}$ (Fig.~\ref{fig:4}(a)(I-IV); also see \href{https://drive.google.com/file/d/1s_I7h_9y7jOZ7cHswhzCyaTbAZvAi8VP/view?usp=drive_link}{movie S2} and \href{https://drive.google.com/file/d/1orVoeLIf3RiVZKojPV0cV-bMAtrdiBCJ/view?usp=drive_link}{movie S3}; in movie S3, the trajectory color coded with $\bar{v}_d$ is stitched and shown in $16\pi$ domain for continuity and clarity, since the microswimmer exits and re-enters the $8\pi$ domain of the simulation, while the concentration field is shown in $8\pi$).
For $\mathrm{Pe}\leq 6$, the microswimmer exhibits steady quasi-ballistic swimming as explained before (Fig.~\ref{fig:4}(a)(I) and (b)(i)).
With increasing viscosity of the swimming medium, i.e., with increasing $\mathrm{Pe}$ beyond 6, the trajectories from DNSs show that the microswimmer exhibits exceedingly frequent reorientations in the swimming direction resulting in the loss of unidirectional swimming (Fig.~\ref{fig:4}(a)(II-IV)).
Furthermore, the reorientations in the swimming direction are accompanied by changes in $\bar{v}_d$. 
Specifically, the local reorientation of the microswimmer is preceded and followed by its slowing down and speeding up respectively (Fig.~\ref{fig:4}(a)(II-IV)).
These predictions closely mirror the experimental observations for increasing $\mathrm{Pe}$ (Fig.~\ref{fig:4}(b)(ii-iii)).

Statistical analyses of the numerical simulation results further substantiate the aforementioned observations made based on the trajectories.
The average swimming speed increases until $\mathrm{Pe} \sim 6$ while the associated standard deviation $std_{v_d}$ decreases, representing steady, quasi-ballistic swimming (Fig.~\ref{fig:4}(c)).
For $\mathrm{Pe} > 6$, the probability distribution $(P(\bar{v}_d))$ of $\bar{v}_d$ becomes progressively broader with increasing $\mathrm{Pe}$ (Fig.~\ref{fig:4}(d)) indicating the increasing fluctuations in the propulsion speed over the trajectory.
With increasing $\mathrm{Pe}$, on the one hand, the increasing probability of events with low swimming speed represents the higher occurrence of the slowing down of the microswimmer prior to reorientation (Fig.~\ref{fig:4}(d)). 
On the other hand, the gradual emergence of the tail of the distribution represents the local acceleration of the microswimmer following the reorientation. 
As a consequence of such fluctuations in the propulsion speed, $\langle {v}_d \rangle$ gradually decays, while the associated standard deviation ${\rm std}_{v_d}$ increases, with increasing $\mathrm{Pe}$ (Fig.~\ref{fig:4}(c)).
Such variation of $\langle{v}_d\rangle$ with $\mathrm{Pe}$ as shown in Fig.~\ref{fig:4}(c), and the corresponding variation in the swimming characteristics about $Pe \sim 6$, are commensurate with results from established numerical framework \cite{hu2019chaotic} (see Appendix, Fig. \ref{fig:supp4} for details).
Further, at high $\Pe$, the average speed follows the scaling $\langle v_d \rangle \sim \Pe^{-1/3}$, in agreement with the prediction of Michelin and Lauga~\cite{michelin2014phoretic} (see Appendix, Fig.~\ref{fig:supp5}).

The apparent loss of ballistic swimming with increasing Pe is also captured by the mean-squared displacement (MSD) plots (Fig.~\ref{fig:4}(e)).
With increasing $\mathrm{Pe}$, the scaling for MSD deviates from the ballistic $(\propto \Delta \bar{t}^2)$ to the sub-ballistic $(\propto \Delta \bar{t}^{1.5})$ at long times $(\Delta \bar{t} \sim \mathcal{O}(1)-\mathcal{O}(10))$.
Furthermore, the loss of unidirectional swimming with increasing $\mathrm{Pe}$ due to frequent reorientations of the microswimmer is evident from the corresponding variations of $C_{vv}$ with time interval $\Delta \bar{t}$ (Fig.~\ref{fig:4}(f)). 
For $\mathrm{Pe}\leq 6$, $C_{vv}$ remains constant about 1 indicating unidirectional swimming. 
However, for $\mathrm{Pe}> 6$, $C_{vv}$ decays with $\Delta \bar{t}$ highlighting the gradual decorrelation between the swimming orientations with time interval. 
Note that beyond $\mathrm{Pe} \sim 35$, the decay in $C_{vv}$ with $\Delta \bar{t}$ becomes relatively slower (Fig.~\ref{fig:4}(f)) indicating that the decorrelation rate itself varies non-monotonically with $\mathrm{Pe}$.    
The slower decay in $C_{vv}$ at higher $\mathrm{Pe}$ stems from the fact that reorientations over shorter time intervals are weaker due to more frequent slowing down of the microswimmer including local stalling. 
However, the normalized velocity autocorrelation function ($VACF (\Delta t) = \left\langle
\frac{\boldsymbol{v_d}(t_0+\Delta t)\cdot\boldsymbol{v_d}(t_0)}
     {\lvert\boldsymbol{v_d}(t_0)\rvert^{2}}
\right\rangle_{t_0} \label{eq:vacf_norm}$); inset Fig.~\ref{fig:4}(f)) shows a monotonic decay with increasing $\mathrm{Pe}$. This deviation in the orientational and velocity correlations at high $\Pe$ arises because the swimming speed decorrelates over short time intervals---due to frequent local slowdowns---before the orientation does.

\subsection{Chemo-hydrodynamic interactions at higher Péclet numbers}
\begin{figure}
    \centerline{\includegraphics[width=1.0\linewidth]{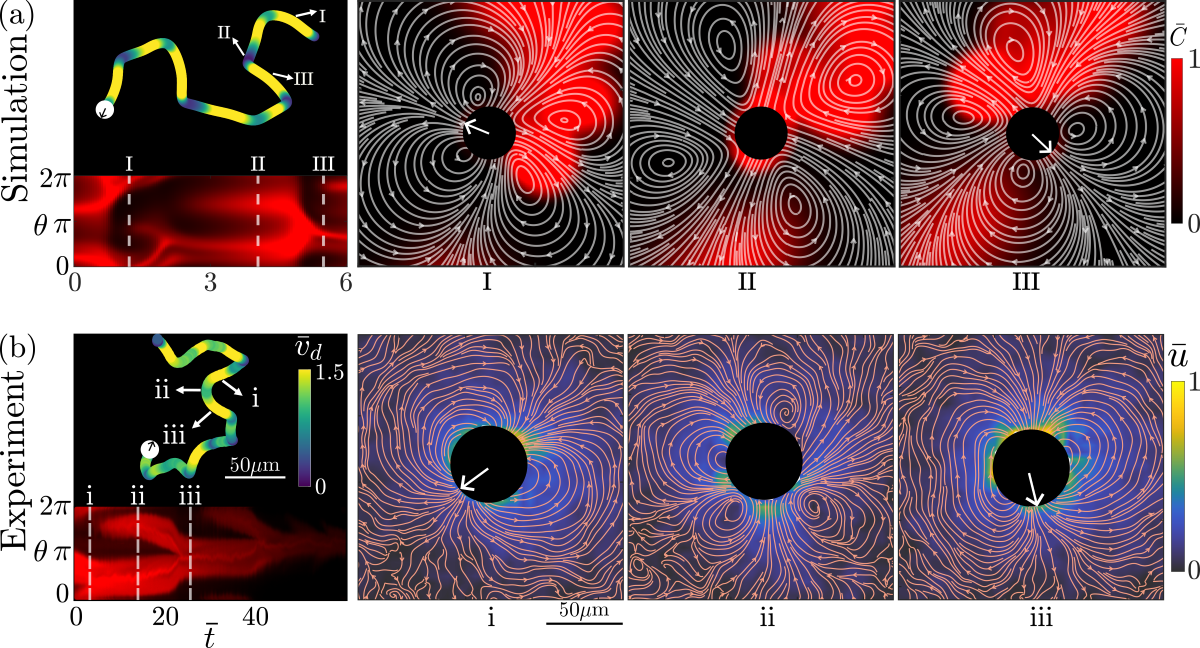}}
    \caption{
\textbf{Chemical and hydrodynamic signatures at relatively large $\mathrm{\bf Pe}$.}
(a) \textit{Top row (Simulation)}: The microswimmer’s trajectory, color-coded by non-dimensionalized propulsion speed $\bar{v}_d$, for $\mathrm{Pe}=58$, and the corresponding kymograph for the chemical concentration around the microswimmer's periphery over a portion of the trajectory. 
We have marked three instants in the trajectory and the kymograph over a typical reorientation event at large $\mathrm{Pe}$-- (I) an instant during which the microswimmer undergoes ballistic swimming prior to a local reorientation in the swimming direction, (II) an instant when the microswimmer is in a state of transient arrest, and (III) an instant when the microswimmer resumes its swimming following a reorientation event.
The flow field (streamlines) generated by the microswimmer along with the corresponding distribution of the chemical concentration at each of the three instants are also shown here.
(b)  \textit{Bottom row (Experiment):} The experimentally obtained trajectory of an active droplet for a comparable $\mathrm{Pe}=50$, along with the corresponding filled micelle kymograph over a part of the trajectory. 
The white disk representing the active droplet is not to scale. 
Here we have also marked three similar instants (i, ii, and iii) over a reorientation event as in the numerically obtained trajectory. 
The flow fields of the active droplet at these three instants are also shown separately using streamlines and the contour plots for the local flow speed $(\bar{u})$.} 
\label{fig:5}
\end{figure}

At higher $\mathrm{Pe}$ the microswimmer transitions from steady motion to disordered, reorientation-dominated trajectories, and it is essential to establish how the proposed numerical framework captures this transition.

With increasing viscosity, and hence increasing $\mathrm{Pe}$, two primary changes occur: (i) the strength of the stresslet contribution to the microswimmer-generated disturbance field increases (Eq.~\ref{eq:stress_gaussian}), and (ii) the diffusive timescale of the ejected chemical becomes longer relative to the advective timescale, since $D(\nu)$ decreases with increasing $\nu$.

 The increasing stresslet strength imparts the quadrupolar structure to the microswimmer flow field during its self-propulsion at higher $\mathrm{Pe}$ (Fig. \ref{fig:5}(a)I), instead of the dipolar flow field at low $\mathrm{Pe}$ (Fig. \ref{fig:2}(c); also see \href{https://drive.google.com/file/d/1ZaVlNjNYODFIMgEs9HTXmQ9Ccdw8wuZy/view?usp=drive_link}{movie S4}).
 When this quadrupolar contribution transiently dominates the flow, the microswimmer is locally slowed down or even stalled while simultaneously pumping out chemical along two diametrically opposite regions on its periphery (Fig.~\ref{fig:5}(a)II). 

 The concentration of the ejected chemical does not undergo a diffusive relaxation, but is locally advected by the flow field since the diffusive time scale is longer than the advective time scale.
 This advection-dominated transport results in two branches of the chemical trail for the locally slowed down/stalled microswommer (instant II in the kymograph in Fig. \ref{fig:5}(a)), instead of the one chemical branch observed during self-propulsion.
 Eventually, the chemical concentration around the microswimmer's periphery develops sufficient non-zero gradient to trigger renewed self-propulsion of the microswimmer along a direction different from its prior swimming orientation, and thereby undergoing a reorientation (compare Figs. \ref{fig:5}(a)III and \ref{fig:5}(a)I).
Although the choice of the new swimming direction is non-deteriminstic in nature, it has to be along a direction where the microswimmer does not interact with its own chemical trail (preserved due to slow diffusion) because of its inherent anti-chemotactic nature.
Note that at the onset of the renewed self-propulsion, the two chemical branches again merge into a single posterior chemical wake (instant III in the kymograph in Fig. \ref{fig:5}(a) and Fig. \ref{fig:5}(a)III).
The swimming continues until the quadrupolar contribution slows it down again, marking the onset of the next reorientation event. 

The fact that the dynamics of the microswimmer at higher $\mathrm{Pe}$ is dominated by the aforementioned reorientation events, which in turn are guided by the microswimmer's tendency to avoid its own slowly diffusing chemical trail (chemotactic avoidance), results in a sub-ballistic, rather than diffusive, MSD scaling $(\propto \Delta \bar{t}^{1.5})$ at long times (Fig.~\ref{fig:4}(e)).
Note that this scaling for MSD is compatible with self-avoiding random walk of active particles; an observation that has also been made experimentally for autophoretic microswimmers, like active droplets, at higher $\mathrm{Pe}$ \cite{hokmabad2021emergence}.  

The observations from the numerical simulations explained above are comparable to our experimental observations for the unsteady chemohydrodynamic characteristics of active droplets over a reorientation event at higher $\mathrm{Pe}$ (Fig. \ref{fig:5}(b)).
The active droplet shows similar dominance of the quadrupolar flow structure during a local arrest in the motion (Fig. \ref{fig:5}(b)ii), along with the generation of two branches for the filled micelle concentration around its periphery (instant ii in kymograph in Fig. \ref{fig:5}(b)).
The new direction of swimming following the reorientation (Fig. \ref{fig:5}(b)iii) is also in accordance with the anti-chemotactic nature of the active droplet. 
One point of difference to be noted here is that the active droplet exhibits a dipolar flow structure (Fig. \ref{fig:5}(b)i and iii) during its self-propulsion instead of a quadrupolar flow field as in the numerical simulation (Fig. \ref{fig:5}(a)I and III).
This is because for the size of the droplet microswimmer used in the experimental system the resulting flow field during self-propulsion is dominated by the dipolar structure; experiments with a relatively smaller active droplet in a quasi-2D setup (but at the same $\mathrm{Pe}$, which could have been achieved by adjusting the viscosity accordingly) would have shown a quadrupolar flow field during its self-propulsion (see supplementary material in \cite{guchhait2025flow}).
 
\subsection{Pairwise interactions and chemotactic avoidance of autophoretic microswimmers}
\begin{figure}
    \centerline{\includegraphics[width=1.0\linewidth]{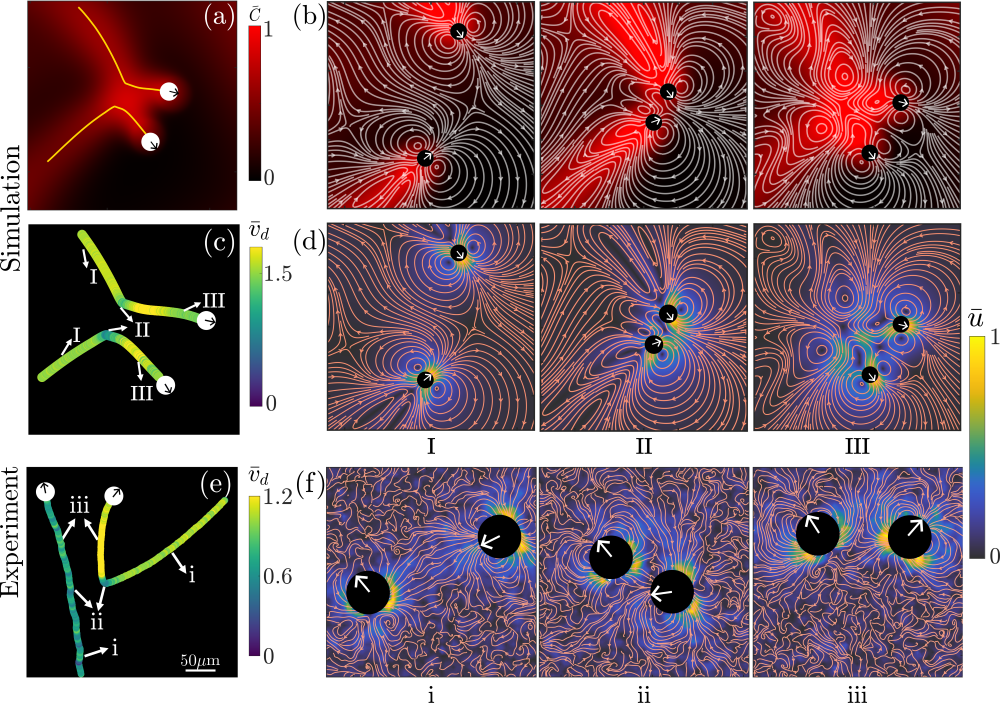}}
\caption{\textbf{Pairwise interaction of active microswimmers at low viscosity, highlighting their chemotactic avoidance.}
\textit{Top two rows (Simulation):} (a) Trajectories of two microswimmers, overlaid on the chemical concentration distribution, as they swim towards one another.
(c) The same trajectories color-coded by the non-dimensional swimming speed $\bar{v}_d$. 
Snapshots showing (b) the streamlines overlaid on the chemical concentration distribution , and (d) the streamlines overlaid on the flow field in terms of the local flow speed $\bar{u}$ at three different instants-- (I) as the two microswimmers approach each other, (II) at the instant of their closest approach, and (III) after the microswimmers move away from each other.
\textit{Bottom row (Experiment):} (e) Experimental results showing the interaction between two active droplets.
(f) Streamlines overlaid on the flow fields taken at corresponding moments (i–iii) show a similar sequence of interactions as observed in the numerical simulations.}
   \label{fig:6}
\end{figure}

In this final subsection, we examine to what extent the numerical framework proposed here can capture the pairwise interaction of two autophoretic microswimmers.
The simulations are performed at $\mathrm{Pe} \approx 4$, with two swimmers of the same radius $a$ swimming in the same domain.
When two microswimmers interact with each other, the steric repulsion in Eq.~\ref{eq:steric} becomes essential, preventing the swimmers from overlapping as they come close. 
Another key difference from the single-swimmer case is that during a pairwise interaction each swimmer can now interact with the hydrodynamic disturbance and the ejected chemical field  emitted by the other. 
This chemo-hydrodynamic interaction significantly alters the swimming dynamics, and the velocities of the microswimmers, which are then updated according to Eq.~\ref{eq:position_update}, capturing the details of the interaction.

Fig.~\ref{fig:6}, rows 1 and 2, show snapshots from the simulations that illustrate how two microswimmers interact with each other using chemical and hydrodynamic cues.
In panels Fig.~\ref{fig:6}(a) and (c), the trajectories of the two swimmers are shown as they approach each other, `sense' the others' chemical trail, and then turn away; also see  \href{https://drive.google.com/file/d/1X2zAyRnKSmiN0aR3ixL5jF56CM8hhpnB/view?usp=sharing}{movie S5}.
In Fig.~\ref{fig:6}(a), the trajectory is overlaid on top of the chemical concentration field, while in Fig.~\ref{fig:6}(c) the trajectory is colour coded with $\bar{v}_d$.
Initially, both swimmers swim following linear trajectories in a quasi-ballistic manner with comparable speeds (Fig.~\ref{fig:6}(c)) as explained before for low Péclet number.
The chemical concentration distribution along with the streamlines, and the flow field during such an instant of approach are shown in Fig.~\ref{fig:6} b(I) and d(I) respectively. 
Note that during such an instant, both microswimmers swim steadily with their chemical wake fixed at the rear apex, diametrically opposite to their swimming direction  (Fig.~\ref{fig:6} b(I)), as also discussed in Fig.~\ref{fig:3}(a).
The microswimmmers also swim with a dipolar flow field (Fig.~\ref{fig:6} d(I)) that is commensurate with the single-swimmer behaviour at the same $\mathrm{Pe}\approx 4$ (Fig.~\ref{fig:2}(c)). 
As the two microswimmers come closer to one another, each starts to interact with the chemical ejected by the other.
At such an instant, both microswimmers slow down (instant II in Fig.~\ref{fig:6}(c), Fig.~\ref{fig:6}(b)II and (d)II), and simultaneously they begin to orient away from each other.
Eventually, each swimmer gradually recovers its ballistic swimming, original swimming speed, and the dipolar flow field as they swim away from each other (instant III in Fig.~\ref{fig:6}(c), Fig.~\ref{fig:6}(b)III and (d)III). 
The reorientation cannot be attributed to long-range hydrodynamic interactions, because at $Pe\approx4$ the swimmers generate predominantly irrotational, rapidly decaying dipolar flows whose rotational influence is weak compared to the chemical cue.
The reorientation of the microswimmers is solely due to the physical interaction of one microswimmer with the chemical wake emanating from the other.
Hence, the numerical framework captures the anti-chemotactic nature of the autophoretic microswimmer leading to chemotactic avoidance. 

The inferences from the numerical simulations as mentioned above are compatible with the observations from our experiments (Fig.~\ref{fig:6} (bottom row)). 
Fig.~\ref{fig:6}(e) shows the trajectories of two active droplets color-coded by  $\bar{v}_d$ during a pair-wise interaction. 
Note that, as observed in the simulations, the active droplet swimming with almost constant speed along a straight path slows down and changes its swimming direction (instant ii) when it encounters the chemical trail left behind by another active droplet.
Eventually, the swimming speed of the active droplet gradually increases as it moves along the new direction, and it resumes quasi-ballistic swimming with a dipolar flow field (instant iii).
Since our numerical framework can capture the chemotactic avoidance of autophoretic microswimmers like active droplets, we think that it will be able to recreate/predict the collective behaviour of such microswimmers stemming from chemo-hydrodynamic ineractions \cite{hokmabad2022chemotactic}. 

\section{Concluding remarks}
\label{sec:conclusions}
Our numerical framework provides a self-consistent description of the dynamics of autophoretic microswimmers, where the surface slip velocity, hydrodynamic signature, chemical concentration field, and particle motion coevolve. The proposed numerical framework has the following three advantageous features--\\
(i) the chemical source is defined at the microswimmer's center and smoothly distributed over the domain using a hyperbolic tangent profile. This lets us evolve the chemical field by solving a single advection–diffusion equation on a fixed grid, without imposing the flux conditions on the moving surface. This avoids explicit moving-boundary treatment and improves numerical stability. Hence, the slip velocity comes from the concentration differences on the particle surface, so interactions between many particles can be captured naturally without approximations.\\
(ii) in the proposed model the disturbance triggered by the microswimmer, i.e. the strength of the stresslet, is not prescribed a priori; instead, both the slip velocity and the stresslet emerge consistently from the evolving  chemical field distribution around the microswimmer. As a result, the strength and direction of the stresslet adjust spontaneously at every time step, capturing the coupled interaction between the chemical transport and the surrounding flow field. \\
(iii) at the same time, the finite size of the microswimmer is included in the modeling framework in a simple way by spreading the computed stresslet smoothly around the particle’s centre using a Gaussian envelope. This method circumvents treating the swimmer as a point-like particle, and allows it to generate a hydrodynamic signature that is realistic even in its near-field features as established through comparison with experiments. The smoothly distributed stress is added as a body force in the momentum equation. The numerical framework then updates the full Navier–Stokes equations with this extra active forcing at every time step.

At low Péclet numbers, the essential features of the swimming dynamics, like swimming trajectory, orientation, and hydrodynamic signature, are in good agreement with the experiments performed with active droplets. 
Quantitative measures such as mean-squared displacement, angular autocorrelation, and local flow speeds further confirm that the simulations accurately capture the swimming characteristics of the autophoretic microswimmers. 
Additionally, the simulations also show that the microswimmer swims steadily by leaving behind a uniform chemical trail which is fixed at the rear apex of the microswimmer, diametrically opposite to its swimming direction.
The spatio-temporal evolution of this self-generated chemical trail in the swimming medium is diffusive in nature, which is consistent with the experimental observations concerning the filled micelle trail left behind by active droplets.

By choosing the diffusion strength of the ejected chemical to vary inversely with the viscosity of the swimming medium, the model consistently captures the corresponding variations in the Péclet number.
As the Péclet number is increased, the swimmer exhibits reduced directional stability and increasingly disturbed motion- a trend which is also observed in our experiments and has been established in the literature \cite{morozov2019nonlinear,hokmabad2021emergence}.
This transition in the swimming dynamics is reflected in the swimming trajectory and in a broader distribution of non-dimensionalised propulsion speed.
The mean-squared displacement transitions from ballistic to sub-ballistic scaling consistent with self-avoiding random walk, indicating a gradual loss of persistent motion. 
A key point to be noted here is that the angular autocorrelation behaves non-monotonically— with increasing Péclet number, the autocorrelation function decays faster as the swimmer reorients more often, but beyond a certain Péclet number the decay becomes slower as the microswimmer stalls more frequently before reorienting.
In contrast, the velocity autocorrelation decreases monotonically with increasing Péclet number denoting a loss of short-time velocity memory. 
Our numerical framework captures the dominance of the quadrupolar flow field with increasing Péclet number and the associated advection-driven transport of the ejected chemical species. 
This feature, coupled with the anti-chemotactic nature of the simulated microswimmer in-built into the framework, remarkably captures the emergent disordered motion of the microswimmer with increasing $\mathrm{Pe}$- a hallmark of the high $\mathrm{Pe}$ regime \cite{hu2019chaotic, hokmabad2021emergence}.

The ability of our simulation to preserve a memory of the microswimmer's trajectory by consistently evolving the chemical trail, along with the anti-chemotactic nature of the simulated microswimmer, help to closely mimic the pair-wise interaction of autophoretic microswimmers as demonstrated in the literature \cite{hokmabad2022chemotactic} and by our experiments using active droplets. 
Specifically, the simulations closely reproduce the trail avoidance behaviour where a microswimmer responds to the chemical field of another microswimmer by steering away from it.
Hence, our numerical simulation can successfully capture chemical-cue-based signalling that could play an important role in collective behaviour.Although the hydrodynamic boundary condition for the autophoretic microswimmer implemented in the numerical framework is the phoretic slip condition, the numerical results closely mimic the experimental observations using active droplets which in reality satisfy a more complex boundary condition involving both Marangoni and phoretic effects.
This further establishes the generality of the proposed numerical modeling technique for autophoretic microswimmers. 

It is important to note certain limitations of the present framework. In its current form, the slip velocity is evaluated solely from the chemically induced (phoretic) contribution. Any additional sources of slip—such as those arising from background flows or other activity mechanisms—are not accounted for. As a result, in situations where slip is not chemically generated, the model predicts zero swimming velocity and, consequently, no hydrodynamic disturbance, which does not reflect the physical behavior of such systems. A second limitation is that sub-particle-scale variations in slip are not resolved. Instead, the surface activity is coarse-grained to yield an effective swimming velocity and an associated stresslet. As a result, a hydrodynamic disturbance appears only when these local contributions produce a non-zero net propulsion. Configurations with zero swimming velocity, therefore, behave as passive particles and do not generate any hydrodynamic disturbance within this framework. In addition, the model employs a simplified representation of activity through $\alpha>0$ or $\alpha<0$, corresponding to extensile (pusher) and contractile (puller) swimmers, respectively, and does not explicitly include neutral swimmers. These limitations can be addressed, depending on the application, by incorporating higher-order hydrodynamic modes beyond the stresslet contribution or by modifying the stresslet formulation.

The current framework provides a versatile platform to explore the dynamics of autophoretic microswimmers at multiple scales. 
For individual swimmers, it allows for systematic variation of key parameters, such as microswimmer size, chemical activity, and swimming strength, over a wide range which may not be trivially accessible by experiments, enabling a detailed understanding of how these factors influence swimming characteristics.
The framework can be extended to simulate the emergent collective behaviour of many interacting microswimmers. Experiments have observed collective effects such as self-induced caging \cite{hokmabad2022chemotactic} and dynamic clustering in systems of chemically active particles \cite{hokmabad2022spontaneously, krueger2016dimensionality}; our numerical framework can be used to simulate and better understand such collective behaviour.
In fact, our numerical model can probe deeper towards investigating how local interactions lead to dynamic clustering, coordinated alignment, large-scale motion, and emergent transport patterns in crowded or confined environments.
By varying parameters like microswimmer density, Péclet number, and chemical interactions, the simulations can provide quantitative predictions of velocity correlations, spatio-temporal organization, and other signatures of emergent collective behavior, complementing and extending experimental insights while opening new avenues for exploring active matter.

\section{Supplementary movies}
\begin{enumerate}
    \item  \href{https://drive.google.com/file/d/1V1NGNPW3XxptnuKSRVPj3R-g8Z1lzTTu/view?usp=drive_link}{movie S1}: Numerical simulations and experiments illustrating the dynamics of the microswimmer at low Péclet number.
    I$^{st}$ quadrant: trajectory of the microswimmer color-coded with instantaneous velocity, II$^{nd}$ quadrant: visualization of the chemical trail as the droplet swims, evaluated using numerical model at $\mathrm{Pe} = 6$;
    III$^{rd}$ quadrant: experimental visualization of the chemical trail, IV$^{th}$ quadrant: experimental trajectory of the microswimmer color-coded with instantaneous velocity at $\mathrm{Pe} = 4$.
    \\
    
    \item \href{https://drive.google.com/file/d/1s_I7h_9y7jOZ7cHswhzCyaTbAZvAi8VP/view?usp=drive_link}{movie S2}: Numerical simulations and experiments illustrating the dynamics of the microswimmer at intermediate Péclet number.
    I$^{st}$ quadrant: trajectory of the microswimmer color-coded with instantaneous velocity, II$^{nd}$ quadrant: visualization of the chemical trail as the droplet swims, evaluated using numerical model at $\mathrm{Pe} = 35$;
    III$^{rd}$ quadrant: experimental visualization of the chemical trail, IV$^{th}$ quadrant: experimental trajectory of the microswimmer color-coded with instantaneous velocity at $\mathrm{Pe} = 15$. 
     \\
    
    \item \href{https://drive.google.com/file/d/1orVoeLIf3RiVZKojPV0cV-bMAtrdiBCJ/view?usp=drive_link}{movie S3}: Numerical simulations and experiments illustrating the dynamics of the microswimmer at the high Péclet number.
    I$^{st}$ quadrant: trajectory of the microswimmer color-coded with instantaneous velocity, II$^{nd}$ quadrant: visualization of the chemical trail as the droplet swims, evaluated using numerical model at $\mathrm{Pe} = 254$;
    III$^{rd}$ quadrant: experimental visualization of the chemical trail, IV$^{th}$ quadrant: experimental trajectory of the microswimmer color-coded with instantaneous velocity at $\mathrm{Pe} = 50$. 
    \\
    
     \item \href{https://drive.google.com/file/d/1ZaVlNjNYODFIMgEs9HTXmQ9Ccdw8wuZy/view?usp=drive_link}{movie S4}: Numerical simulations and experiments illustrating the change in the hydrodynamic signature of the microswimmer at high Péclet number. Left panel: Streamlines (flow field) around the microswimmer plotted on top of the chemical trail, using the numerical model at $\mathrm{Pe} = 58$. Right panel: Streamlines (flow field) around the microswimmer plotted on top of the magnitude of the generated flow field using experiments at $\mathrm{Pe} = 50$.
     \\
     
     \item \href{https://drive.google.com/file/d/1X2zAyRnKSmiN0aR3ixL5jF56CM8hhpnB/view?usp=sharing}{movie S5}: Numerical simulations and experiments illustrating the chemotactic avoidance of the microswimmer at low Péclet number.
    I$^{st}$ quadrant: trajectory of the microswimmers color-coded with instantaneous velocity, II$^{nd}$ quadrant: visualization of change in the direction of swimming of the microswimmer due to chemotactic signal from another microswimmer, evaluated using the numerical model at $\mathrm{Pe} = 4$;
    III$^{rd}$ quadrant: experimental visualization of change in the direction of swimming due to chemotactic signal from another microswimmer, IV$^{th}$ quadrant: experimental trajectory of the microswimmers, color-coded with instantaneous velocity at $\mathrm{Pe} = 4$. 
     \\
     \\
     All movies are playing at 25fps.
     The movies are compiled using ImageJ \cite{schneider2012nih}. 
\end{enumerate}

\section*{Acknowledgments}
These simulations were performed on the Paramseva supercomputers through the National Supercomputing Mission and the IITH Kanad clusters. A.D.R. and A.K. acknowledge the Ministry of Education, Govt. of India (MoE) and Indian Institute of Technology Hyderabad. S.S.S. acknowledges the Prime Minister’s Research Fellowship (PMRF), a scheme by the Govt. of India to improve the quality of research in various research institutions in the country. 
R.D. acknowledges support from Science and Engineering Research Board (SERB) (now subsumed under Anusandhan National Research Foundation), Department of Science and Technology (DST), Government of India, through Grant No. SRG/2021/000892, from IIT Hyderabad through seed Grant No. SG 93, and from Indian Council of Medical Research (ICMR) through Grant No. IIRP-2023-2832. A.G. acknowledges SERB-DST (India) Projects MTR/2022/000232, CRG/2023/007056-G, DST (India) grant no. DST/NSM/R\&D HPC Applications/2021/05 and grant no. SR/FST/PSI-215/2016, and IITH for Seed Grant No. IITH/2020/09.
\\
\textbf{Competing Interests:} The authors declare none.\\
\textbf{ORCID ID:} 
\begin{itemize}
    \item Anupriya Dutta Roy: https://orcid.org/0009-0003-0654-456X
    \item Smita S Sontakke: https://orcid.org/0009-0005-2484-7411
    \item Arvind Kumar: https://orcid.org/0009-0000-9992-1827
    \item Ranabir Dey: https://orcid.org/0000-0002-0514-7357
    \item Anupam Gupta: https://orcid.org/0000-0002-7335-0584
\end{itemize}
\clearpage
\newpage
\appendix
\section*{Appendix}
\label{appendix}

\renewcommand{\thefigure}{A\arabic{figure}}
\setcounter{figure}{0}
\renewcommand{\theHfigure}{A\arabic{figure}}

\section{Validation of the numerical framework}
\subsection{Effect of solute activity layer thickness on trajectory and average swimming speed}

\noindent
The solute activity layer (Eq.~\ref{eq:activity}) has a thickness thickness $\delta \approx 5\Delta x$, which is adequately resolved at the chosen grid spacing (see \S~\ref{sec:numerical}). This choice provides a balance between numerical stability and physical resolution. In particular, a very small value of $\delta$, (e.g., $\delta \approx \Delta x$) produces a sharp concentration gradient that is under-resolved on the grid, leading to numerical instabilities. On the other hand, a large value of $\delta$ (e.g., $\delta \approx 10\Delta x$) makes the activity layer comparable to the swimmer size ($a \sim 40\Delta x$ for our simulations), which is physically inappropriate.
To assess the sensitivity of our results to this parameter, we perform additional simulations with $\delta \approx 3\Delta x$ and $\delta \approx 7\Delta x$. These simulations are carried out at two representative Péclet numbers, $\mathrm{Pe} \approx 4$ and $\mathrm{Pe} \approx 60$, with identical initial particle positions $(0,0)$ and over the same simulation duration. The grid resolution was kept fixed across all cases, consistent with the main simulations. 
We find that variations in the activity layer thickness about the baseline choice $\delta = 5\Delta x$ do not affect the qualitative nature of the particle trajectories, while the corresponding average swimming speed shows only negligible changes (variations $< 1\%$).

\begin{figure}[h]
    \centering
    \includegraphics[width=18cm]{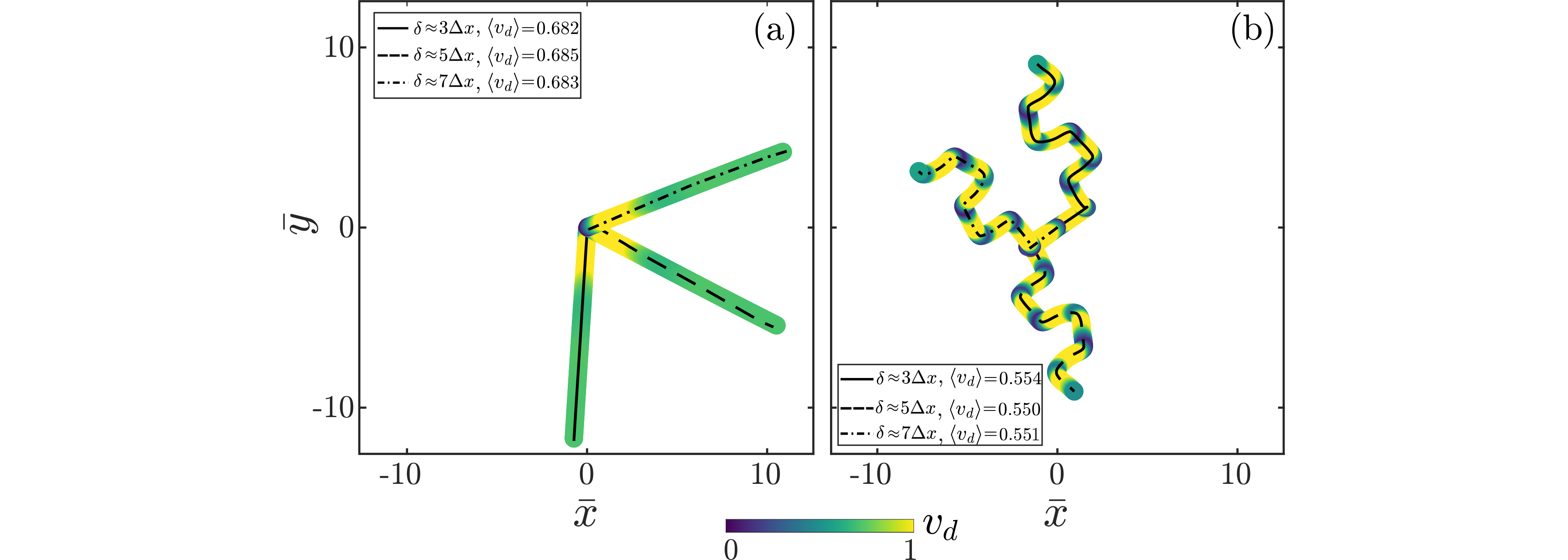}
    \caption{Trajectories and mean swimming speeds for different solute activity layer thicknesses, $\delta \simeq 3\Delta x$, $5\Delta x$, and $7\Delta x$, all resolved at the same grid spacing, at $Pe \approx 4$ (a) and $Pe \approx 60$ (b).}
    \label{figure:supp1}
\end{figure}

\subsection{Radial variation of solute concentration with far-field saturation}

\begin{figure}[h]
    \centering
    \includegraphics[width=18cm]{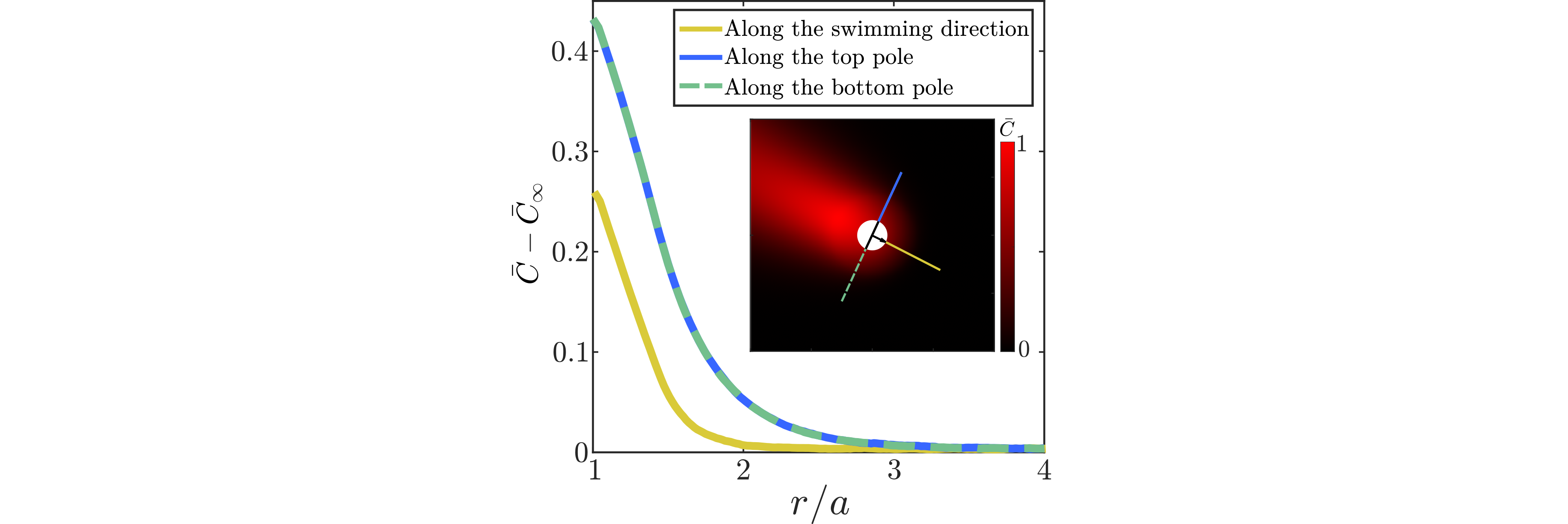}
    \caption{Radial variation of the normalized concentration difference, $\bar{C} - \bar{C}_\infty$, as a function of $r/a$, measured from the particle surface along the swimming (front) and transverse (upper and lower pole) directions. The concentration decays to the far-field value, starting from a lower value along the swimming direction and a higher value along the pole directions, with similar diffusive decay in all directions. The overlap of the pole profiles indicates transverse symmetry. Inset: two-dimensional concentration field around the swimmer.
}
    \label{fig:supp2}
\end{figure}

\noindent
To check the spatial distribution of the solute field, we examine the radial variation of the normalized concentration difference, $\bar{C} - \bar{C}_\infty$, as a function of the distance from the particle surface (see \S.~\ref{sec:conc_emission}). Profiles are extracted along three directions: the swimming direction and two orthogonal directions corresponding to the upper and lower poles of the particle.
Fig.~\ref{fig:supp2} shows these profiles as a function of the normalized distance $r/a$. The concentration decreases smoothly from its maximum value at the particle surface and approaches a constant far-field value (i.e. $\bar{C}-\bar{C}_{\infty}$=0)~\cite{michelin2014phoretic}. We observe a clear directional dependence in the concentration field. Along the swimming direction, the concentration appears lower, as this region is continuously exposed to freshly generated solute. In contrast, along the perpendicular (pole) directions, the concentration remains higher due to the relative accumulation of solute. As a result, the decay along the lateral directions starts from a higher value, while along the forward direction it starts from a relatively lower value. The rate of decay in space is otherwise similar in all directions, which is expected since the solute transport is primarily governed by diffusive decay at low $\mathrm{Pe}$. The close overlap between the upper and lower pole profiles confirms the expected symmetry of the system about the axis perpendicular to the swimming direction. The inset in Fig.~\ref{fig:supp2} shows the corresponding two-dimensional concentration field. 

\newpage
\subsection{Effect of domain size on the critical Péclet number for swimming onset}
\begin{figure}[h]
    \centering
    \includegraphics[width=18cm]{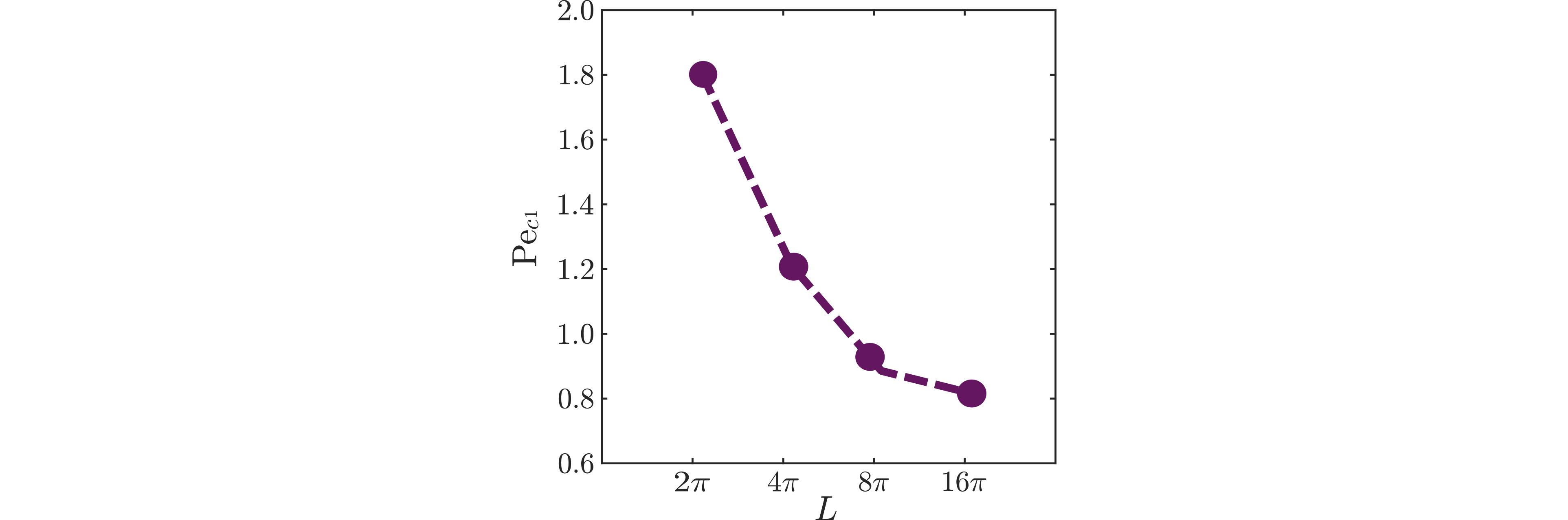}
    \caption{\label{fig:supp3}
Critical Péclet number, $\Pe_{c1}$, for the onset of the swimming mode as a function of the domain size, as obtained for our numerical framework.
}

\end{figure}

\noindent
We investigate the influence of domain size on the onset of the swimming instability by measuring the critical Péclet number, $\Pe_{c1}$, for different box sizes. Fig.~\ref{fig:supp3} shows the variation of $\Pe_{c1}$ as a function of the box size. It is observed that $\Pe_{c1}$ decreases as the box size increases. This indicates that in smaller domains, a higher Péclet number is required to initiate motion, while in larger domains the swimmer starts moving at lower $\Pe$. This behavior can be understood from the dimensional dependence of the concentration field. In three dimensions, the solute concentration decays as $1/r$, so the far-field contribution becomes negligible, and the onset of instability remains independent of the domain size. In contrast, in two dimensions, the concentration decays logarithmically, leading to a slow decay and a strong dependence on the domain size. As a result, the concentration field retains a strong dependence on the system size, which affects the gradients responsible for propulsion. In smaller domains, this leads to weakened gradients and a delayed onset of self-propulsion. In larger domains, this effect is reduced, and the swimmer becomes unstable at lower $\mathrm{Pe}$, consistent with the results of Hu \textit{et al.}~\cite{hu2019chaotic}. In all our simulations, we fix the box size to $8\pi$ (see \S.~\ref{sec:numerical}), which provides a moderately large domain that ensures adequate resolution of concentration gradients while keeping the computational cost tractable.
\newpage
\subsection{Péclet number vs swimming speed}
\begin{figure}[h]
   \centering
    \includegraphics[width=18cm]{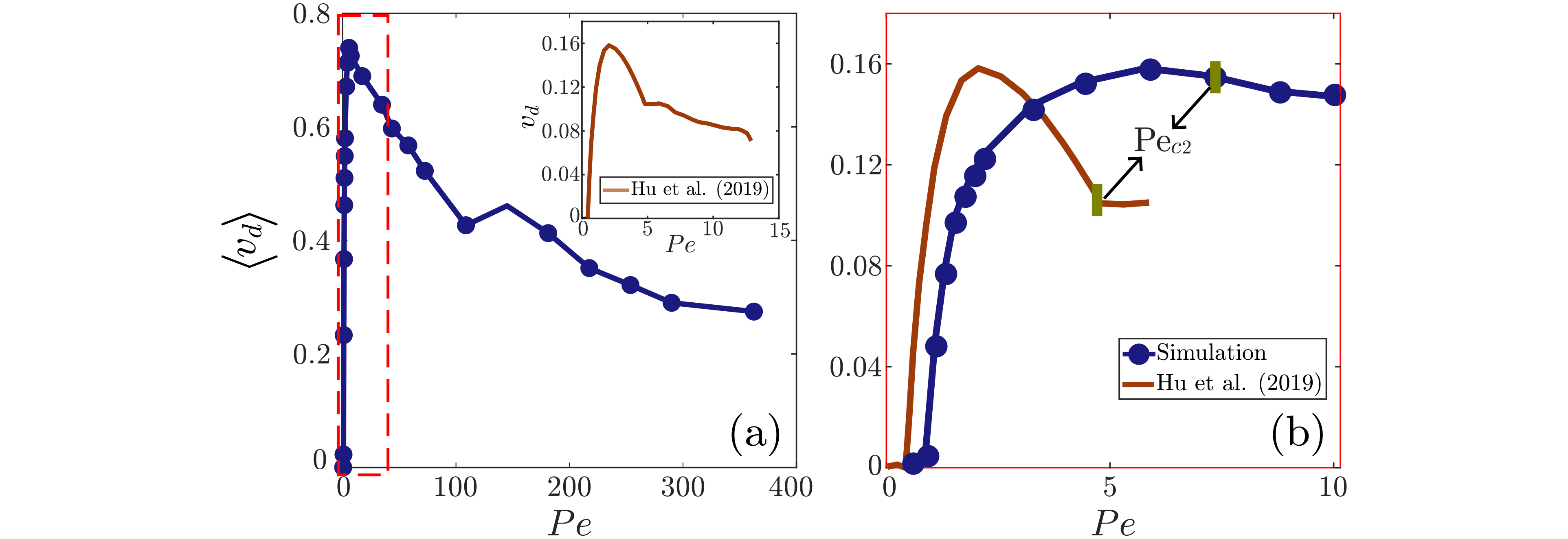}
   \caption{
(a) Average swimming speed as a function of the Péclet number, $\Pe$. The swimmer remains stationary at low $\Pe$, then starts moving, reaches a maximum speed, and decreases at higher $\Pe$. Inset: results from Hu \textit{et al.}~\cite{hu2019chaotic}. (b) Comparison of the average speed from simulations with Hu \textit{et al.}~\cite{hu2019chaotic}. The vertical olive line marks the second critical Péclet number ($\Pe_{c2}$), beyond which the swimmer shows a meandering type of motion.
}
    \label{fig:supp4}
\end{figure}

We examine the variation of the average swimming speed as a function of the $\Pe$ to support the results shown in Fig.~\ref{fig:4}. Figure~\ref{fig:supp4}(a) shows the average speed over a wide range of $\Pe$. At low $\Pe$, the swimmer remains stationary. Beyond a critical value $\Pe_{c1}$, it undergoes a transition to self-propulsion, after which the speed increases, reaches a maximum, and subsequently decreases with further increase in $\Pe$. The inset in Fig.~\ref{fig:supp4} shows the corresponding results reported by Hu \textit{et al.}~\cite{hu2019chaotic}. We note that the definition of the Péclet number differs between their work and our simulations, leading to differences in the numerical values, although the overall trend remains consistent. 
Figure~\ref{fig:supp4}(b) shows a zoomed-in view of the low-$\Pe$ regime up to the second critical Péclet number, $\Pe_{c2}$. This value marks the transition beyond which the swimmer no longer exhibits steady unidirectional motion but instead follows a meandering trajectory, as also observed in our simulations. The value of $\Pe_{c2}$ is indicated by a vertical olive line in the figure. In this regime, the average speed begins to decrease. For quantitative comparison, the simulation data are rescaled using the values reported by Hu \textit{et al.}~\cite{hu2019chaotic}. 

\begin{figure}[h]
    \centering
    \includegraphics[width=18cm]{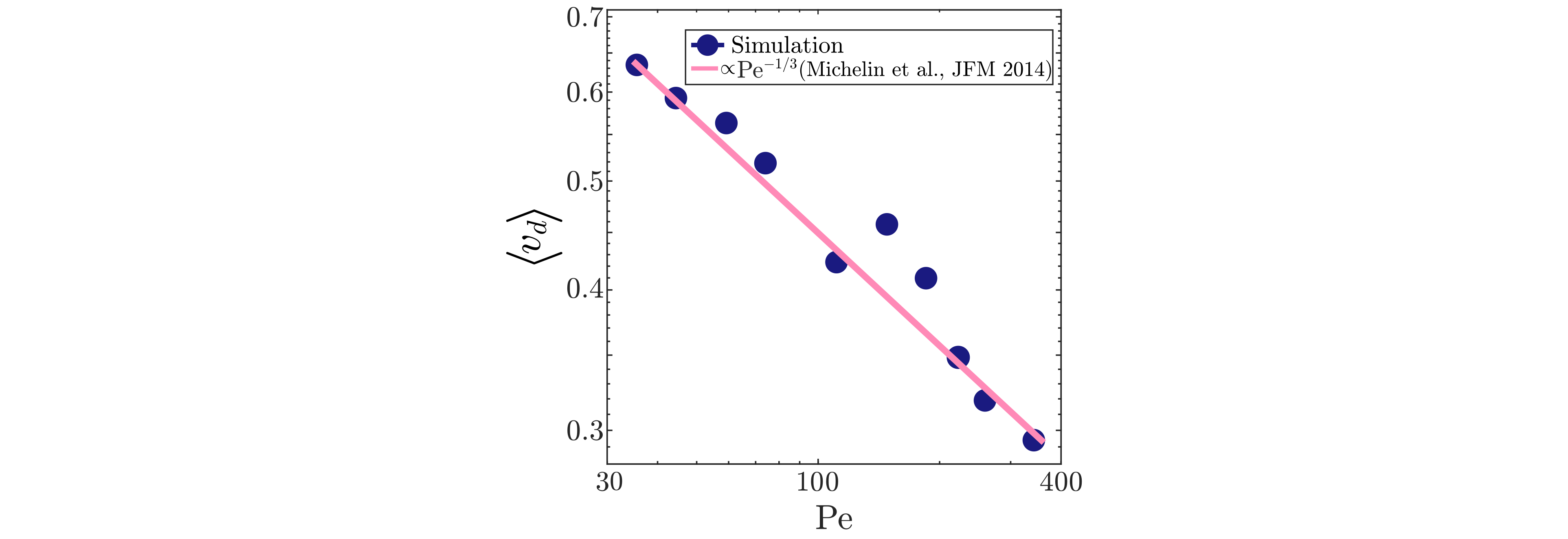}
    \caption{
 Log--log plot of the average speed at higher $\Pe$, showing a decay close to $\langle v_d \rangle \sim \Pe^{-1/3}$ consistent with the observation reported by Michelin and Lauga ~\cite{michelin2014phoretic}}
    \label{fig:supp5}
\end{figure}

Figure~\ref{fig:supp5} shows a close view of the higher-$\Pe$ regime, highlighting the monotonic decrease in the average swimming speed with increasing $\Pe$. 
Over the range $\Pe \approx 35$ to $300$, the simulation results are consistent with the scaling $\langle v_d \rangle \sim \Pe^{-1/3}$ reported by Michelin and Lauga~\cite{michelin2014phoretic}.

\clearpage
\newpage
\bibliography{bibliography}

\end{document}